 \documentclass[12pt, draftclsnofoot, onecolumn]{IEEEtran}
\usepackage[english]{babel}
\usepackage{amsthm}
\usepackage{amsmath,amssymb}
\usepackage{amsfonts}
\usepackage{mathtools}
\usepackage{graphicx}
\usepackage{color}
\usepackage{optidef}
\usepackage{hyperref}
\usepackage{enumitem}
\usepackage{subcaption}
\usepackage{cite}
\usepackage{nccmath}
\usepackage{diagbox}
\usepackage{multirow}
\usepackage{cases} 
\makeatletter
\newif\if@restonecol
\makeatother
\usepackage[linesnumbered,ruled,vlined]{algorithm2e}
\usepackage{algpseudocode}
\DeclareMathOperator*{\minimize}{minimize}

\DeclareMathOperator\erf{erf}
\newtheorem{theorem}{Theorem}

\newtheorem{lemma}[theorem]{Lemma}
\begin{document}
\title{Robust Symbol Level Precoding for Overlay Cognitive Radio Networks} \author{Lu~Liu,~\IEEEmembership{Student Member,~IEEE,}
Christos~Masouros,~\IEEEmembership{Senior Member,~IEEE,}\\
        and~A.~Lee~Swindlehurst,~\IEEEmembership{Fellow,~IEEE}
\thanks{L. Liu and A. Swindlehurst are with Center for Pervasive Communications and Computing, University of California, Irvine, USA (e-mail: \{liul22,swindle\}@uci.edu).}
\thanks{C. Masouros is with the Department of Electronic \& Electrical Engineering, University College London, London WC1E7JE,
U.K. (e-mail: c.masouros@ucl.ac.uk). }
\thanks{This work was supported by the National Science Foundation under
grants CCF-2008714 and CCF-2225575.}}
\maketitle
\begin{abstract}
This paper focuses on designing robust symbol-level precoding (SLP) in an overlay cognitive radio (CR) network, where the primary and secondary networks transmit signals concurrently. When the primary base station (PBS) shares data and perfect channel state information (CSI) with the cognitive base station (CBS), we derive an SLP approach that minimizes the CR transmission power and satisfies symbol-wise Safety Margin (SM) constraints of both primary users (PUs) and cognitive users (CUs). {\color{black} The resulting optimization has a quadratic objective and linear inequality (LI) constraints, which can be solved by standard convex methods.} For the case of imperfect CSI from the PBS, we propose robust SLP schemes. First, with a norm-bounded CSI error model to approximate the uncertain channels, we adopt a max-min philosophy to conservatively achieve robust SLP constraints. Second, we use the additive quantization noise model (AQNM) to describe the quantized PBS CSI and employ a stochastic constraint to formulate the problem. {\color{black} Both robust approaches also result in a quadratic objective with LI constraints. Simulation results show that, rather than simply trying to eliminate the network’s cross-interference, the proposed robust SLP schemes enable the primary and secondary networks to aid each other in meeting their quality of service constraints.}
\end{abstract}

\section{Introduction}
As the number of wireless devices and their applications grow exponentially, the availability of unoccupied radio spectrum is becoming increasingly scarce and occupied bands are increasingly congested. Over the past two decades, cognitive radio (CR) technology has been extensively studied as a means to alleviate this problem through more efficient, flexible and comprehensive use of the spectrum {\color{black}\cite{CRpropose1, CRbreak, ahmad20205g}}. 

The fundamental challenge lies in balancing the interference generated by the CR at the primary users (PUs) with the quality of service (QoS) of the cognitive users (CUs). To address this issue, both the inter-system and inter-user interference need to be successfully managed. In the standard (non-cognitive) multiuser downlink scenario, beamforming or precoding at the multi-antenna transmitter can be employed to mitigate the multiuser interference (MUI) and compensate for its adverse affect on the received signals \cite{intpre}. Existing precoding schemes can be classified as either block-level precoding (BLP) or symbol-level precoding (SLP). In recent decades, many approaches have been proposed to implement block-level precoders that only depend on the current channel state information (CSI), such as maximum ratio transmission (MRT), zero-forcing (ZF), regularized ZF and optimum interference-constrained or power-constrained precoding \cite{MRT,ZF,RZF,optbeam,optEE,sinrcons}. These approaches all treat the MUI as a detrimental effect that is to be suppressed as much as possible.

Unlike BLP, SLP techniques exploit information about the symbols to be transmitted in addition to the CSI, which can significantly improve performance at the expense of increased complexity at the transmitter {\color{black}\cite{mahaSLPsurvey,SLPtutorial2020, salem2021error}}. The additional degrees of freedom (DoF) provided by the symbol-level information make it possible to exploit the constructive component of the MUI, converting it into constructive interference (CI) that can move the received signals further from the decision thresholds of the constellation points \cite{CIZF,masouros2010correlation,Masouros13}. CI-based SLP recasts the traditional viewpoint of interference as a source of degradation to one where interference is a potential resource that can be exploited. 

Constructive interference regions (CIRs), which define the degree to which the received symbols will be robust to noise and unmodeled perturbations, are fundamental to SLP designs. While early CI-based SLP approaches were intended to increase the distance of the CIRs from the symbol decision boundaries, they did not directly optimize the CIR. More recent techniques have focused on designing the precoder to directly optimize this distance \cite{masouros2015exploiting,kalantari2016directional,alodeh2016energy,swindlehurst2017ICASSP,swindlehurst2018reduced,jedda2018quantized,li2018interference,ali2018CIR,ali2018DPCIR,li2020interference}, which has been referred to as the {\em safety margin} (SM). Optimal Maximum Safety Margin (MSM) precoders generally result in a non-linear mapping between the symbols and the transmitted waveforms, and can be shown to minimize an upper bound on the symbol error rate (SER) \cite{jedda2018quantized}. This is contrasted with algorithms that minimize the mean squared-error (MMSE) between the transmitted and received symbols, which do not offer the same guarantee \cite{Masouros13,masouros2015exploiting}. MSM precoders are in general able to achieve a better QoS for the same level of transmit power, or equivalently the same QoS with less power consumption. There has been limited work that studies CI-based SLP in CR systems. Although the linear precoder proposed in \cite{khan2012interference} for an overlay CR network is based on the use of CI with the MMSE criterion, the PU performance is impaired compared to the primary-only case, which violates the principle of CR design that we follow. 

The performance of both BLP and SLP are sensitive to channel uncertainties due for example to channel estimation errors, quantization noise or latency-related effects {\color{black}\cite{jindal2006mimo,vucic2008robust,liang2022power}}. To mitigate the impact of such errors, robust designs are needed that properly model the errors and account for their effect in the optimization of the precoders. \textcolor{black}{Two general approaches for doing so include assuming worst-case bounded error models or exploiting known statistical properties of the CSI error. The former case involves the use of deterministic CSI error bounds that assume the error is confined to a convex uncertainty region (typically an ellipsoid) surrounding the true CSI \cite{pascual2005robust}. In these approaches, robustness is achieved by constraining the users' QoS or other design objectives to be satisfied for all channel realizations in the convex uncertainty region, effectively minimizing the impact of the worst-case channel within the given error bound \cite{vucic2008robust, zheng2009robust}. This max-min philosophy can lead to a relatively conservative design depending on the tightness of {\em a priori} error bound. In the second case, a particular distribution (e.g., Gaussian) is assumed for the error, and Bayesian or other probabilistic approaches \cite{pascual2005robust} are employed to optimize the quality of service (QoS) or transmit power under certain stochastic signal-to-interference-plus-noise-ratio (SINR) or rate-outage probability constraints \cite{zhang2008statistically,shenouda2008probabilistically}. In this case, the probability-constraint formulation is typically not deterministic and \textcolor{black}{various techniques} must be used to obtain a tractable problem \cite{bertsimas2006tractable,ben2009safe}. In either of the two cases described above, the penalty paid for increasing the robustness to imperfect CSI is increased transmit power.}

In overlay or cooperative CR systems, CSI errors beyond those due to channel estimation are anticipated due to the limited cooperation between the PBS and CBS. While robust BLP designs for traditional MIMO or CR scenarios have been widely investigated \cite{zheng2009robust,zheng2010robust,ma2012chance,wajid2013robust}, robust SLP algorithms for general CR scenarios have not been considered. Prior work on robust designs for SLP includes \cite{masouros2015exploiting}, which derived a robust SLP algorithm suitable for imperfect CSI with bounded CSI errors, but it is based on a multicast formulation without fully taking advantage of CI. The work described in \cite{haqiqatnejad2020robust} considered a linear channel distortion model with bounded additive noise and Gaussian-distributed channel uncertainties. They designed robust SLP schemes to minimize transmission power subject to CI constraints as well as QoS or SINR requirements. While not focused on CR applications, this prior work demonstrates that robust SLP designs can be formulated to improve and achieve a better balance between QoS and power consumption. 

In this paper, we propose robust CR SLP algorithms for each of two different CSI error models that account for the quantization error in the CSI shared by the PBS with the CBS. In particular, we focus on overlay CR downlink channels {\color{black}\cite{CRbreak,liang2017cooperative}} where the PBS shares with the CBS its CSI to the PUs and CUs, as well as its data intended for the PUs. The shared CSI is assumed to be quantized, which is known to often make achieving the desired user QoS constraints infeasible without introducing robustness into the problem formulation \cite{huang2010cooperative,suraweera2010capacity}. {\color{black} In addition, the imperfect CSI also means that the PBS precoding is not precisely known at the CBS, and thus the CBS has an imperfect estimate of the transmitted PBS signal, even if the PBS data symbols are perfectly known. This makes finding a robust solution in the cognitive radio case more complicated than in prior SLP-related work, where the transmitted signals are assumed to be perfectly known. If left unaddressed, the combination of these effects will almost certainly} cause the noise-free received symbols at both the PUs and CUs to fall outside the desired CIR. To derive a robust SLP formulation for CR systems, we formulate the problem as one of minimizing the transmit power at the CBS while simultaneously satisfying the SM constraints at both the PUs and CUs to guarantee the worst-case user's QoS. 

{\color{black}
We first derive a power-minimizing SLP approach for overlay CR with SM constraints at both the PUs and CUs assuming perfect CSI, leading to a quadratic optimization problem with linear constraints that can be efficiently solved. We then derive the SM at each user for two different imperfect CSI models, including the effect of the imprecisely known PBS transmit signal. We first consider the case where the quantization error is norm-bounded as in \cite{pascual2005robust}, and we derive a robust SLP algorithm based on maximizing the worst case SM. This leads to a conservative design that trades transmit power for increased protection of the PUs from the CR interference due to the quantized CSI. Then we study a stochastic approach based on the additive quantization noise model (AQNM) \cite{bai2013optimization,bai2015energy} that is sufficiently accurate to approximate the quantization error at low and medium signal-to-noise ratios and has been widely used in the analysis of quantized MIMO systems \cite{fletcher2007robust,fan2015uplink,zhang2017performance}. In this case, the SM of the PUs and CUs are constrained to meet a preset threshold with a certain probability. We then apply the {\em Safe Approximation I} method in \cite{haqiqatnejad2020robust} to reformulate the intractable probabilistic constraints as deterministic constraints and finally construct an optimization problem to obtain the robust SLP solution.

The use of SLP for overlay CR has not been considered previously in the literature. The work in \cite{khan2012interference} is the most related prior effort, but it requires that the CBS directly transmits the PBS data together with its own data, which is not as energy efficient as our proposed approach. In addition, unlike our proposed approaches, \cite{khan2012interference} does not consider the impact of the PBS interference at the cognitive users, it does not assume imprecise knowledge of the PBS waveform, it uses a less effective SLP technique, and it does not take into account the fact that the PBS CSI exploited at the CBS may be imperfect due to quantization or other effects. Most notably, our proposed SLP algorithms enable the PUs to exploit constructive interference as well as the CUs, and thus we can demonstrate that the presence of the cognitive network can actually {\em improve} the PU network performance rather than degrade it. This result is unique to the literature on CR, which focuses on not impairing the PU QoS.

We conduct a number of simulations assuming the PBS channel is quantized using the scalar Lloyd Max algorithm that minimizes the average quantization noise power \cite{max1960quantizing,lloyd1982least}. These simulations demonstrate the flexibility of the proposed robust SLP algorithms in trading transmit power for improved performance when quantized CSI is present. They further demonstrate the ability of the proposed methods to improve the performance of both the primary and cogntive networks. Furthermore, inspired by the results of \cite{yang2018limited}, we study the problem of allocating bits to the CSI of the PBS to the PUs and CUs, and demonstrate that the bit allocation strategy in our robust SLP algorithm is not as important as that in the non-robust methods. Note that a subset of the results presented in this paper were previously reported in \cite{lu2023overlay}.}

\textsl{Notation:} Bold lower case and upper case letters indicate vectors and matrices, and non-bold letters express scalars. The $N\times N$ identity (zero) matrix is denoted by $\mathbf{I}_{N}\: (\mathbf{0}_{N\times N})$. The $N$ dimensional vector of ones (zeroes) is denoted by $\mathbf{1}_{N} \: (\mathbf{0}_{N} )$. $\mathbf{A}_{mn}$ denotes the $(m,n)$-th element in the matrix $\mathbf{A}$ and $a_{m}$ denotes the $m$-th element in the vector $\mathbf{a}$. The operators $(\cdot)^{*}$, $(\cdot)^{-1}$, $(\cdot)^{\text{T}}$ and $(\cdot)^{\text{H}}$ stand for the conjugation, the inverse, the transpose and the Hermitian transpose operations, respectively. $\mathbb{C}^{m\times n}$ represents the space of complex matrices of dimension $m\times n$. $\mathbb{E}(\cdot)$, $\lvert(\cdot)\rvert$ and $\|\cdot\|$ respectively represent the expectation operator, absolute value and the Euclidean norm. $\mathcal{CN}(\mu, \sigma^2)$ denotes the complex normal distribution with mean $\mu$ and variance $\sigma^2$. The functions $\text{tr}\{\cdot\}$ and $\text{diag}\{\cdot\}$ respectively indicate the trace of a matrix and a vector composed of the diagonal elements of a square matrix, while $\text{diag}\{\mathbf{a}\}$ denotes a square diagonal matrix with the elements of vector $\mathbf{a}$ on the main diagonal. $\mathcal{R}\{\cdot\}$ and $\mathcal{I}\{\cdot\}$ denote the real and imaginary parts of a complex number, respectively.  For matrices and vectors, $\geq$ and $\leq$ denote element-wise inequalities,  {\color{black} and $\otimes$ denotes the Kronecker product.} 
\begin{table}[h!]
\color{black}
\centering
\caption{Definition for all acronyms}
\begin{tabular}{|c|c|} 
\hline
\textbf{Acronym} & \textbf{Full Name} \\
 \hline
SLP&Symbol Level Precoding\\
 \hline
CR&Cognitive Radio\\ 
 \hline
PBS&Primary Base Station\\
 \hline
PU&Primary User\\
 \hline
CBS&Cognitive Base Station\\
 \hline
CU&Cognitive User\\
 \hline
CSI&Channel State Information\\
 \hline
SM&Safety Margin\\
 \hline
LI&Linear Inequality\\
 \hline
AQNM&Additive Quantization Noise Model\\
 \hline
 QoS & Quality of Service \\
\hline
MUI & Multi-User Interference \\
\hline
BLP & Block-Level Precoding \\
\hline
SLP & Symbol-Level Precoding \\
\hline
MRT & Maximum Ratio Transmission \\
\hline
ZF & Zero-Forcing \\
\hline
DoF & Degrees of Freedom \\
\hline
CI & Constructive Interference \\
\hline
CIR & Constructive Interference Region\\
\hline
MSM & Maximum Safety Margin \\
\hline
SER & Symbol Error Rate \\
\hline
MMSE & Minimum Mean Squared-Error \\
\hline
SINR & Signal-to-Interference-plus-Noise Ratio \\
\hline
AWGN & Additive White Gaussian Noise \\
\hline
PALP & Phase Alignment Linear Precoding\\
\hline
SR & Symbol Region \\
\hline
CIR & Constructive Interference Region \\
\hline
PMSLP & Power-Minimizing SLP \\
\hline
BLER & Block Error Rate \\
\hline
EE & Energy Efficiency \\
 \hline
\end{tabular}
\end{table}

\section{System Model and Problem Formulation}\label{secsystemmodel}
\begin{figure}[!t]
\centering
\includegraphics[width=2.5in]{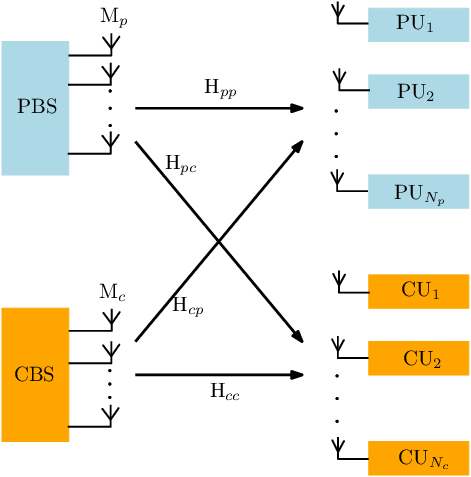}
\caption{Cognitive Radio System Model}
\label{systemmodel}
\end{figure}

We consider a downlink CR network with an $M_c$-antenna CBS serving $N_c$ single-antenna CUs. The CR network is granted access to share the primary system spectrum in which an $M_p$-antenna PBS is communicating with $N_p$ single-antenna PUs. The system model is depicted in Fig.~\ref{systemmodel}. The direct primary and cognitive channels are assumed to be respectively denoted by the following flat-fading model: 
\begin{eqnarray}
\mathbf{H}_{pp} & = & \begin{bmatrix}
\mathbf{h}_{pp,1}^{\text{T}}&\cdots&\mathbf{h}_{pp,N_p}
^{\text{T}}\end{bmatrix}^{\text{T}}\in \mathbb{C}^{N_p\times M_p} \\ 
\mathbf{H}_{cc} & = & \begin{bmatrix}
\mathbf{h}_{cc,1}^{\text{T}}&\cdots&\mathbf{h}_{cc,N_p}
^{\text{T}}\end{bmatrix}^{\text{T}}\in \mathbb{C}^{N_c\times M_c} 
\end{eqnarray}
The corresponding interference channels are defined as 
\begin{eqnarray}
\mathbf{H}_{pc} & = & \begin{bmatrix}
\mathbf{h}_{pc,1}^{\text{T}}&\cdots&\mathbf{h}_{pc,N_p}
^{\text{T}}\end{bmatrix}^{\text{T}}\in \mathbb{C}^{N_c\times M_p} \\ 
\mathbf{H}_{cp} & = & \begin{bmatrix}
\mathbf{h}_{cp,1}^{\text{T}}&\cdots&\mathbf{h}_{cp,N_p}
^{\text{T}}\end{bmatrix}^{\text{T}}\in \mathbb{C}^{N_p\times M_c}
\end{eqnarray}
from the PBS to CUs and the CBS to PUs, respectively. We will leave further specification of the channel models until later. 

The vectors $\mathbf{s}_p(t)=[s_{p,1}(t),s_{p,2}(t),\cdots,s_{p,N_p}(t)]^T$ and $\mathbf{s}_c(t)=[s_{c,1}(t),s_{c,2}(t),\cdots,s_{c,N_c}(t)]^T$ will be used to represent the symbols to be transmitted to the individual PUs and CUs, respectively, at time $t$. In this work we assume for simplicity that all transmitted symbols are uncorrelated and drawn from a $D$-PSK constellation with unit magnitude, i.e., $s_{l,m}(t)\in\{s|s={\color{black} \exp(j\pi(2d+1)/D)},\ d\in\{0,\cdots, D-1\}\}$ where $l\in \{p,c\}$ denotes the primary or cognitive system, and $m$ denotes the user index in the corresponding system. The sets $\mathcal{K}=\{1,\cdots,N_p\}$ and $\mathcal{J}=\{1,\cdots,N_c\}$ enumerate the PUs and CUs, respectively. The idea of CI precoding can in principle be applied to any constellation design \cite{ali2018DPCIR}, e.g., QAM \cite{jedda2018quantized} otherwise, but is most easily formulated for the case of PSK signals. {\color{black} The algorithm for other constellations such as QAM is slightly more complicated since the definition of safety margin becomes dependent on whether an inner, edge, or corner constellation point is transmitted, but the basic principle of the algorithm is the same.}

At time slot $t$, the received signals at the PUs and CUs can be respectively written as
\begin{eqnarray}
\mathbf{y}_p(t)=\mathbf{H}_{pp}\mathbf{x}_p(t)+\mathbf{H}_{cp}\mathbf{x}_c(t)+\mathbf{n}_p(t)\label{PUsymbol}\\
\mathbf{y}_c(t)=\mathbf{H}_{cc}\mathbf{x}_c(t)+\mathbf{H}_{pc}\mathbf{x}_p(t)+\mathbf{n}_c(t)\label{CUsymbol}
\end{eqnarray}
where $\mathbf{x}_p(t)\in \mathbb{C}^{M_p\times 1}$ and $\mathbf{x}_c(t)\in \mathbb{C}^{M_c\times 1}$ are the transmitted signals at the PBS and CBS after precoding and power loading, and $\mathbf{n}_p(t)\sim \mathcal{CN}(0,\sigma_p^2) $ and $\mathbf{n}_c(t) \sim \mathcal{CN}(0,\sigma_c^2)$ are additive white Gaussian noise (AWGN) vectors. In order to simplify the notation, in what follows we drop the time index $t$. 

\subsection{{\color{black} Phase Alignment Linear Precoding (PALP)}}\label{PALPexpression}
\begin{figure}[!t]
\centering
\includegraphics[width=2.5in]{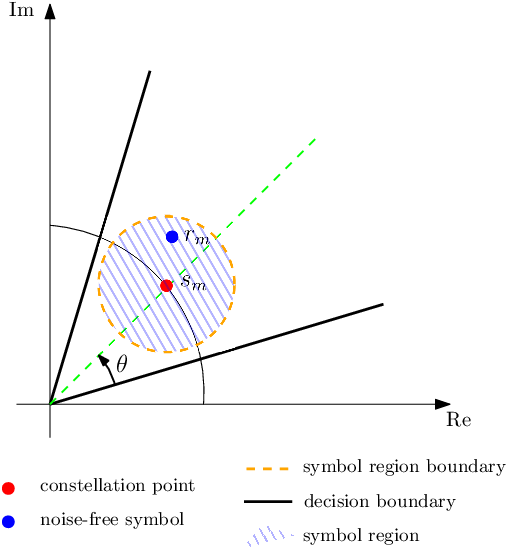}
\caption{Symbol region for conventional precoding}
\label{condition}
\end{figure}

Conventional precoding methods such as MMSE, ZF and maximum-SINR beamforming are designed with the objective of minimizing the inter-user interference so that the received symbols lie as close as possible to the nominal constellation points (or scaled versions thereof in the case of PSK). 
This is effectively equivalent to ensuring that for each user $m$, the noise-free received signal $r_m=\mathbf{h}_m\mathbf{x}$ lies within a circle centered at its corresponding constellation point $s_m$ \cite{masouros2015exploiting}, as depicted in Fig.~\ref{condition}. The shaded area inside the circle is referred to as the symbol region (SR), a down-scaled version of the decision region for $s_m$.

The method described in \cite{khan2012interference} is based on the MMSE criterion and the early {\color{black} Phase Alignment Linear Precoding (PALP)} technique for SLP \cite{masouros2010correlation}; it is the prior approach most closely related to the algorithms we present in this paper for cognitive radio scenarios. However, a modification to the PALP approach in \cite{khan2012interference} is necessary for a fair comparison, and to allow the algorithm to protect the PUs from the CR interference. In particular, we tailor \cite{khan2012interference} (hereafter referred to as CR-PALP) by allowing different instantaneous power scaling factors at the PBS and CBS:
\begin{gather}
f_p=\sqrt{\frac{P_p}{\text{trace}\{\mathbf{W}_p\mathbf{s}_p\mathbf{s}_p^H\mathbf{W}_p^H\}}},\
f_c=\sqrt{\frac{P_c}{\text{trace}\{\mathbf{W}_c\mathbf{s}_c\mathbf{s}_c^H\mathbf{W}_c^H\}}}
\end{gather}  
where {\color{black} $P_p$ and $P_c$ respectively denote the total transmit power of the PBS and CBS, $f_p$ and $f_c$ are the respective instantaneous scaling factors, and $\mathbf{W}_p$ and $\mathbf{W}_c$ are the linear precoders for the primary and cogntive systems, respectively.} The generalized MSE criterion for CR-PALP is given by
\begin{equation}
\epsilon=\mathbb{E}\{\|\mathbf{V}_p\mathbf{s}+\mathbf{HW}_c\mathbf{s}-(\mathbf{A}+\mathbf{B}\mathbf{Q}^{\phi})\mathbf{s}\|^2\}
\end{equation}
where according to \cite{khan2012interference},
{\color{black}
\[
\mathbf{V}_p=\begin{bmatrix}\mathbf{H}_{pp}\mathbf{W}_p&\mathbf{0}_{N_p\times N_c}\\ \mathbf{H}_{pc}\mathbf{W}_p&\mathbf{0}_{N_c\times N_c}\end{bmatrix} \; , \quad \mathbf{s}=\begin{bmatrix}\mathbf{s}_p\\\mathbf{s}_c\end{bmatrix} \; , \quad \mathbf{H}=\begin{bmatrix}
    \mathbf{H}_{cp}\\\mathbf{H}_{cc}
\end{bmatrix} \; ,
\] }
$\mathbf{A}=\text{diag}\{[1, \cdots, 1, 0, \cdots, 0]\}$ is a diagonal matrix whose first $N_p$ diagonal elements equal~$1$, $\mathbf{B}=\text{diag}\{[0, \cdots, 0, 1, \cdots, 1]\}$ is a diagonal matrix whose last $N_c$ elements equal~$1$, and $\mathbf{Q}^{\phi}=\text{diag}(\mathbf{s})\cdot |\mathbf{HH}^H| \cdot \text{diag}(\mathbf{s})^H$ contains the phase-corrected correlation elements. The precoding matrix at the CBS derived from the MMSE criterion is given by
\begin{equation}
\mathbf{W}_{c}=\mathbf{H}^H(\mathbf{HH}^H)^{-1}(\mathbf{A}+\mathbf{B}\mathbf{Q}^{\phi}-\mathbf{V}_p)\; , 
\end{equation}
and the received signals at the PUs and CUs are
\begin{gather}
\mathbf{y}_p=f_p\mathbf{s}_p+\mathbf{n}_{p}\\
\mathbf{y}_c=(f_p-f_c)\mathbf{H}_{pc}\mathbf{W}_p\mathbf{s}_p+f_c\mathbf{Q}^{\phi}_c\mathbf{s}_c+\mathbf{n}_{c}\; .
\end{gather}
The performance of the PUs in our CR scenario will not be impaired using this modified CR-PALP approach, unlike using the method of \cite{khan2012interference} directly.

\subsection{SM-constrained SLP}\label{secsm}
\begin{figure}[!t]
\centering
\includegraphics[width=2.5in]{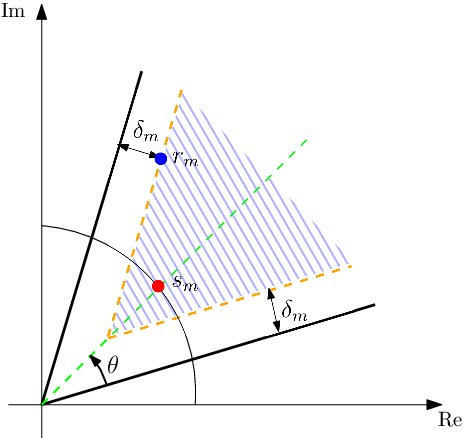}
\caption{Symbol region for CI-Based SLP}
\label{SM1}
\end{figure}

\begin{figure}[!t]
\centering
\includegraphics[width=2.5in]{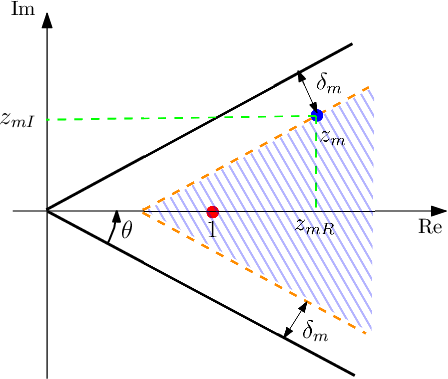}
\caption{Symbol region and safety margin in a modified coordinate ststem}
\label{SM2}
\end{figure}
For PSK constellations, it is not necessary that $r_m$ be close to $s_m$ in order to be decoded correctly, as long as it lies in the correct decision region with a given level of certainty. Thus, it is not necessary that all of the inter-user interference be eliminated, since some interference components could add constructively and push the received symbol further into the decision region, making it more robust to noise and interference external to the system. We can thus redefine the SR as, for example, in Fig.~\ref{SM1}, where the SR becomes a displaced version of the circular sector of angular extent $2\pi/D$ centered at the origin and corresponding to $s_m$. This displaced sector has an infinite radius, and all points within it are at least a certain distance $\delta_m$ from the decision boundaries for $s_m$. This region is referred to as a constructive interference region (CIR) with safety margin (SM) $\delta_m$ \cite{jedda2017massive}. The larger $\delta_m$, the more robust the received signal will be to noise, interference, modeling errors, or other impairments.

In order to mathematically interpret the CIR and SM in a unified way, we rotate the original coordinate system by the negative phase of the desired constellation symbol, i.e., $\measuredangle s_m^*$, to obtain the modified coordinate system in Fig.~\ref{SM2}. After rotation, $s_m$ is placed at $1$ on the real axis, and $r_m$ is relocated to
\begin{equation}
    z_m=r_ms_m^* \; .
\end{equation}
Then we can easily calculate the SM of the noise-free symbol at user $m$ as {\color{black} \cite{jedda2017massive,jedda2018quantized}}
\begin{equation}
    \delta_m=\mathcal{R}\{z_m\}\sin\theta-|\mathcal{I}\{z_m\}|\cos\theta\label{safetymarginorigin}.
\end{equation}
Ideally, the SM should be large enough to sufficiently reduce the probability that noise or other impairments will push the noise-free signal outside the desired detection region; the larger the SM, the smaller the SER. To design the precoder, one can constrain the SM to be above a certain threshold to ensure a given target SER. The fact that the CIR in Fig.~\ref{SM2} is much larger than the SR in Fig.~\ref{secsm} means that increased flexibility is available to achieve the given performance objective. In this paper, we will consider the following type of SLP optimization, which minimizes the transmit power to achieve a certain desired SM:
\begin{align}
\min_{\mathbf{x}}\quad &\lVert\mathbf{x}\rVert^2
\label{MSM}\\
\text{subject to}\quad &\delta_m\geq\delta_{m,0}\quad \forall m\in \mathcal{M}
\end{align}
where $\delta_{m,0}$ is the desired minimum SM for user $m$ and $\mathcal{M}=\{1,\cdots,M\}$ indexes the users. 

\section{Power Minimization SLP in CR}\label{perfect}
Before considering the robust SLP design, we first examine the simpler case where the PBS shares its data and perfect CSI with the CBS. \textcolor{black}{The SM for each PU and CU is assumed to be constrained to be $\delta_{p,k}^0$ for $k \in \mathcal{K}$ and $\delta_{c,j}^0$ for $j\in \mathcal{J}$, corresponding for example to possibly different target SERs for each PU and SU}. Here we focus on SLP designs that minimize the transmit power at the CBS and achieve the SM QoS constraints at both the PUs and CUs.

\subsection{Primary System}
The rotated symbols received at the PUs can be expressed as
\begin{equation}
    z_{p,k}=s_{p,k}^*r_{p,k}=s_{p,k}^*(\mathbf{h}_{pp,k}\mathbf{x}_p+\mathbf{h}_{cp,k}\mathbf{x}_c)\label{zpk}
\end{equation}
for $k\in\mathcal{K}$. Defining
\begin{equation}
\tilde{\mathbf{h}}_{pp,k}\triangleq s_{p,k}^*\mathbf{h}_{pp,k},\quad
\tilde{\mathbf{h}}_{cp,k}\triangleq s_{p,k}^*\mathbf{h}_{cp,k}\label{tildep} \; ,
\end{equation}
we have
\begin{equation}z_{p,k}=\tilde{\mathbf{h}}_{pp,k}\mathbf{x}_p+\tilde{\mathbf{h}}_{cp,k}\mathbf{x}_c.
\end{equation}
The constraints ensuring the QoS of the PUs can be expressed as
\begin{eqnarray}
\delta_{p,k}=\mathcal{R}\{z_{p,k}\}\sin \theta-\lvert\mathcal{I}\{z_{p,k}\}\rvert\cos\theta\geq{\color{black}\delta_{p,k}^0} \,, \forall k\in\mathcal{K},\label{perfectPcons}
\end{eqnarray}
which is equivalent to 
\begin{subnumcases}{}
\mathcal{R}\{\tilde{\mathbf{h}}_{pp,k}\mathbf{x}_p\}\sin \theta-\mathcal{I}\{\tilde{\mathbf{h}}_{pp,k}\mathbf{x}_p\}\cos\theta\nonumber\\\quad+\mathcal{R}\{\tilde{\mathbf{h}}_{cp,k}\mathbf{x}_c\}\sin \theta-\mathcal{I}\{\tilde{\mathbf{h}}_{cp,k}\mathbf{x}_c\}\cos\theta\geq \color{black}{\delta_{p,k}^0}&{}\\
\mathcal{R}\{\tilde{\mathbf{h}}_{pp,k}\mathbf{x}_p\}\sin \theta+\mathcal{I}\{\tilde{\mathbf{h}}_{pp,k}\mathbf{x}_p\}\cos\theta\nonumber\\\quad+\mathcal{R}\{\tilde{\mathbf{h}}_{cp,k}\mathbf{x}_c\}\sin \theta+\mathcal{I}\{\tilde{\mathbf{h}}_{cp,k}\mathbf{x}_c\}\cos\theta\geq\color{black}{\delta_{p,k}^0}&{}\;.
\end{subnumcases}

For any given complex vector $\mathbf{x}$, we define the operator
\begin{equation}
\mathrm{\mho}(\mathbf{x})\triangleq \begin{bmatrix}
\mathcal{R}\{\mathbf{x}\}\sin\theta-\mathcal{I}\{\mathbf{x}\}\cos\theta&-\mathcal{R}\{\mathbf{x}\}\cos\theta-\mathcal{I}\{\mathbf{x}\}\sin\theta\\
\mathcal{R}\{\mathbf{x}\}\sin\theta+\mathcal{I}\{\mathbf{x}\}\cos\theta&\mathcal{R}\{\mathbf{x}\}\cos\theta-\mathcal{I}\{\mathbf{x}\}\sin\theta
\end{bmatrix}
\label{operator}
\end{equation}
and denote
\begin{equation}
\widetilde{\mathbf{H}}_{pp,k}^{\mathrm{\mho}}=\mathrm{\mho}(\tilde{\mathbf{h}}_{pp,k}),\quad
\widetilde{\mathbf{H}}_{cp,k}^{\mathrm{\mho}}=\mathrm{\mho}(\tilde{\mathbf{h}}_{cp,k}).
\end{equation}
Using the following real-valued notation,
\begin{equation}
\check{\mathbf{x}}_p=\begin{bmatrix}\mathcal{R}\{\mathbf{x}_p\}\\\mathcal{I}\{\mathbf{x}_p\}\end{bmatrix},\quad \check{\mathbf{x}}_c=\begin{bmatrix}\mathcal{R}\{\mathbf{x}_c\}\\\mathcal{I}\{\mathbf{x}_c\}\end{bmatrix},
\end{equation}
the constraints in Eq.~(\ref{perfectPcons}) can be simplified as
\begin{equation}
\widetilde{\mathbf{H}}_{pp,k}^{\mathrm{\mho}}\check{\mathbf{x}}_p+\widetilde{\mathbf{H}}_{cp,k}^{\mathrm{\mho}}\check{\mathbf{x}}_c\geq{\color{black}\delta_{p,k}^0}\mathbf{1}_2 \quad \forall k\in\mathcal{K} \; .
\end{equation}

\subsection{Cognitive System}
Similarly, the rotated symbols at the CUs can be written as
\begin{eqnarray}
z_{c,j}=s_{c,j}^*r_{c,j} & = & s_{c,j}^*(\mathbf{h}_{cc,j}\mathbf{x}_c+\mathbf{h}_{pc,j}\mathbf{x}_p) \\
& = & \tilde{\mathbf{h}}_{cc,j}\mathbf{x}_c+\tilde{\mathbf{h}}_{pc,j}\mathbf{x}_p \; ,
\end{eqnarray}
where
\begin{equation}
\tilde{\mathbf{h}}_{cc,j}\triangleq s_{c,j}^*\mathbf{h}_{cc,j}\quad
\tilde{\mathbf{h}}_{pc,j}\triangleq s_{c,j}^*\mathbf{h}_{pc,j}\label{tildec}\; .
\end{equation}
The SM constraints at the CUs can be expressed as
\begin{subnumcases}{}
\mathcal{R}\{\tilde{\mathbf{h}}_{cc,j}\mathbf{x}_c\}\sin \theta-\mathcal{I}\{\tilde{\mathbf{h}}_{cc,j}\mathbf{x}_c\} \cos\theta\nonumber\\\quad+\mathcal{R}\{\tilde{\mathbf{h}}_{pc,j}\mathbf{x}_p\}\sin \theta-\mathcal{I}\{\tilde{\mathbf{h}}_{pc,j}\mathbf{x}_p\} \cos\theta\geq\color{black}{\delta_{c,j}^0}&{}\\
\mathcal{R}\{\tilde{\mathbf{h}}_{cc,j}\mathbf{x}_c\}\sin \theta+\mathcal{I}\{\tilde{\mathbf{h}}_{cc,j}\mathbf{x}_c\}\cos\theta\nonumber\\\quad+\mathcal{R}\{\tilde{\mathbf{h}}_{pc,j}\mathbf{x}_p\}\sin \theta+\mathcal{I}\{\tilde{\mathbf{h}}_{pc,j}\mathbf{x}_p\}\cos\theta\geq \color{black}{\delta_{c,j}^0}&{}
\end{subnumcases}
for $j\in\mathcal{J}$, which can again be written more compactly using the operator in Eq.~(\ref{operator}):
\begin{equation}
\widetilde{\mathbf{H}}_{cc,j}^{\mathrm{\mho}}\check{\mathbf{x}}_c+\widetilde{\mathbf{H}}_{pc,j}^{\mathrm{\mho}}\check{\mathbf{x}}_p\geq{\color{black}\delta_{c,j}^0}\mathbf{1}_2 \quad \forall j\in\mathcal{J} \; .
\end{equation}

Combining all of the above notation together, we can express the general power minimization SLP problem
with perfect CSI as follows:
\begin{eqnarray}
\minimize_{\check{\mathbf{x}}_c} &\|\check{\mathbf{x}}_c\|^2\tag{P1}\label{P1-1}\\
\text{subject to}&
\begin{bmatrix}
-\widetilde{\mathbf{H}}_{cp}^{\mathrm{\mho}}\\-\widetilde{\mathbf{H}}_{cc}^{\mathrm{\mho}}
\end{bmatrix}\check{\mathbf{x}}_c\leq\begin{bmatrix}
\widetilde{\mathbf{H}}_{pp}^{\mathrm{\mho}}\\
\widetilde{\mathbf{H}}_{pc}^{\mathrm{\mho}}
\end{bmatrix}\check{\mathbf{x}}_p-\begin{bmatrix}
\color{black}
\boldsymbol{\delta}_{p}^0\otimes\mathbf{1}_{2}\\
\color{black}\boldsymbol{\delta}_{c}^0\otimes\mathbf{1}_{2}
\end{bmatrix}\tag{C1}\label{C1}
\end{eqnarray}
{\color{black}where the inequalities are to be interpreted element-wise,  $\boldsymbol{\delta}_{p}^0=\begin{bmatrix}
\delta_{p,1}^0&\cdots&\delta_{p,N_p}^0\end{bmatrix}^T$, $\boldsymbol{\delta}_{c}^0=\begin{bmatrix}
\delta_{c,1}^0&\cdots&\delta_{c,N_c}^0\end{bmatrix}^T$, and $\widetilde{\mathbf{H}}_{ab}^{\mathrm{\mho}}\triangleq\mathrm{\mho}(\text{diag}(\mathbf{s}_b^*)\mathbf{H}_{ab})$, with $a,b \in \{c,p\}$.}  The result is a quadratic programming problem with linear inequality constraints which can be efficiently solved using a variety of numerical methods.

\section{Robust SLP for Norm-Bounded CSI Errors}\label{NormBound}
In practice, the CSI shared by the PBS with the CBS will be imperfect due for example to quantization, or somewhat outdated due to delays required for processing and transmission. As a result, robust precoding designs are critical for overlay systems. We address such a design in this section for the case where the imperfect CSI can be described in terms of a norm-bounded error. We model the CSI shared by the PBS with the CBS as follows:
\begin{gather}
\mathbf{h}_{pp,k}=\hat{\mathbf{h}}_{pp,k}+\mathbf{e}_{p,k}\\
\mathbf{h}_{pc,j}=\hat{\mathbf{h}}_{pc,j}+\mathbf{e}_{c,j},
\end{gather}
where the $\hat{\cdot}$ indicates the shared CSI and $\mathbf{e}_{p,k}, \mathbf{e}_{c,j}$ are norm-bounded CSI error vectors, i.e., $\|\mathbf{e}_{p,k}\|_2\leq \epsilon_{p,k}$ and $\|\mathbf{e}_{c,j}\|_2\leq \epsilon_{c,j}$. No other assumption regarding the channels is required. Using Eq.~(\ref{operator}), it is easy to show that
\begin{align}{}
\widetilde{\mathbf{H}}_{pp,k}^{\mathrm{\mho}}=\mathrm{\mho}(\tilde{\mathbf{h}}_{pp,k})&=\mathrm{\mho}(s_{p,k}^*(\hat{\mathbf{h}}_{pp,k}+\mathbf{e}_{p,k}))\\
&=\bar{\mathbf{H}}_{pp,k}^{\mathrm{\mho}}+\tilde{\mathbf{E}}_{p,k}^{\mathrm{\mho}} \\
\widetilde{\mathbf{H}}_{pc,j}^{\mathrm{\mho}}=\mathrm{\mho}(\tilde{\mathbf{h}}_{pc,j})&=\mathrm{\mho}(s_{c,j}^*(\hat{\mathbf{h}}_{pc,j}+\mathbf{e}_{c,j}))\\
&=\bar{\mathbf{H}}_{pc,j}^{\mathrm{\mho}}+\tilde{\mathbf{E}}_{c,j}^{\mathrm{\mho}},
\end{align}
{\color{black}where $\bar{\mathbf{h}}_{pp,k}\triangleq s_{p,k}^*\hat{\mathbf{h}}_{pp,k}$, $\bar{\mathbf{H}}_{pp,k}^{\mathrm{\mho}}\triangleq \mathrm{\mho}(\bar{\mathbf{h}}_{pp,k})$, $\tilde{\mathbf{E}}_{p,k}^{\mathrm{\mho}}\triangleq \mathrm{\mho}(s_{p,k}^*\mathbf{e}_{p,k})$, $\bar{\mathbf{h}}_{pc,j}\triangleq s_{c,j}^*\hat{\mathbf{h}}_{pc,j}$, $\bar{\mathbf{H}}_{pc,j}^{\mathrm{\mho}}\triangleq \mathrm{\mho}(\bar{\mathbf{h}}_{pc,j})$, and $\tilde{\mathbf{E}}_{c,j}^{\mathrm{\mho}}\triangleq \mathrm{\mho}(s_{c,j}^*\mathbf{e}_{c,j})$. Due to the uncertainty in $\mathbf{h}_{pp,k}$, the transmitted signal at the PBS, i.e., $\mathbf{x}_p$, which necessarily depends on $\mathbf{h}_{pp,k}$, is not precisely known. Assuming that the precoding method used at the PBS is known to the CBS, we will assume that an estimate of the transmitted signal, denoted by $\check{\mathbf{x}}_p^e$, can be computed by the CBS using the quantized CSI $\hat{\mathbf{h}}_{pp,k}$.  With this notation,} the constraints in \ref{P1-1} can be reformulated as 
\begin{gather}
(\bar{\mathbf{H}}_{pp,k}^{\mathrm{\mho}}+\tilde{\mathbf{E}}_{p,k}^{\mathrm{\mho}}){\color{black}\check{\mathbf{x}}_p^e}+\widetilde{\mathbf{H}}_{cp,k}^{\mathrm{\mho}}\check{\mathbf{x}}_c\geq
{\color{black}\delta_{p,k}^0}\mathbf{1}_2,\; \forall k\in\mathcal{K}\\
\widetilde{\mathbf{H}}_{cc,j}^{\mathrm{\mho}}\check{\mathbf{x}}_c+(\bar{\mathbf{H}}_{pc,j}^{\mathrm{\mho}}+\tilde{\mathbf{E}}_{c,j}^{\mathrm{\mho}}){\color{black}\check{\mathbf{x}}_p^e}\geq{\color{black}\delta_{c,j}^0}\mathbf{1}_2,\; \forall j\in\mathcal{J} .
\end{gather}

For a robust bounded-CSI-error design, we desire that the above constraints hold for every possible error realization and every user:
\begin{eqnarray}
-\tilde{\mathbf{E}}_{p,k}^{\mathrm{\mho}}{\color{black}\check{\mathbf{x}}_p^e}\leq \bar{\mathbf{H}}_{pp,k}^{\mathrm{\mho}}{\color{black}\check{\mathbf{x}}_p^e}+\widetilde{\mathbf{H}}_{cp,k}^{\mathrm{\mho}}\check{\mathbf{x}}_c-{\color{black}\delta_{p,k}^0}\mathbf{1}_2\; , \nonumber\\
\forall \|\mathbf{e}_{p,k}\|_2\leq \epsilon_{p,k},\; \forall k\in\mathcal{K}\; ,\label{normbound1}\\
-\tilde{\mathbf{E}}_{c,j}^{\mathrm{\mho}}{\color{black}\check{\mathbf{x}}_p^e}\leq\widetilde{\mathbf{H}}_{cc,j}^{\mathrm{\mho}}\check{\mathbf{x}}_c+\bar{\mathbf{H}}_{pc,j}^{\mathrm{\mho}}{\color{black}\check{\mathbf{x}}_p^e}-{\color{black}\delta_{c,j}^0}\mathbf{1}_2\; ,\nonumber\\
\forall \|\mathbf{e}_{c,j}\|_2\leq \epsilon_{c,j},\;\forall j\in\mathcal{J}\; .\label{normbound2}
\end{eqnarray}
We separate the operator $\mathrm{\mho}(\mathbf{x})$ into two parts, as follows: 
\begin{equation}
\mathrm{\mho}(\mathbf{x})\triangleq \begin{bmatrix}
\mathrm{\mho}_1(\mathbf{x})\\
\mathrm{\mho}_2(\mathbf{x})
\end{bmatrix}\triangleq\begin{bmatrix}
\mathbf{x}^{\mathrm{\mho}_1}\\
\mathbf{x}^{\mathrm{\mho}_2}
\end{bmatrix},
\end{equation}
where
\begin{eqnarray}
\mathbf{x}^{\mathrm{\mho}_1}&=&\begin{bmatrix}
\mathcal{R}\{\mathbf{x}\}\sin\theta-\mathcal{I}\{\mathbf{x}\}\cos\theta&-\mathcal{R}\{\mathbf{x}\}\cos\theta-\mathcal{I}\{\mathbf{x}\}\sin\theta
\end{bmatrix}\nonumber\\
&=&\mathrm{\mho}_1(\mathbf{x}),
\end{eqnarray}
\begin{eqnarray}
\mathbf{x}^{\mathrm{\mho}_2}&=&\begin{bmatrix}
\mathcal{R}\{\mathbf{x}\}\sin\theta+\mathcal{I}\{\mathbf{x}\}\cos\theta&\mathcal{R}\{\mathbf{x}\}\cos\theta-\mathcal{I}\{\mathbf{x}\}\sin\theta
\end{bmatrix}\nonumber\\
&=&\mathrm{\mho}_2(\mathbf{x}),
\end{eqnarray}
so that constraint (\ref{normbound1}) can be rewritten in two parts as
\begin{eqnarray}
-\tilde{\mathbf{e}}_{p,k}^{\mathrm{\mho}_1}{\color{black}\check{\mathbf{x}}_p^e}\leq \bar{\mathbf{h}}_{pp,k}^{\mathrm{\mho}_1}{\color{black}\check{\mathbf{x}}_p^e}+\widetilde{\mathbf{h}}_{cp,k}^{\mathrm{\mho}_1}\check{\mathbf{x}}_c-{\color{black}\delta_{p,k}^0}\; , \quad
\forall \|\mathbf{e}_{p,k}\|_2\leq \epsilon_{p,k},\; \forall k\in\mathcal{K}\; ,\\
-\tilde{\mathbf{e}}_{p,k}^{\mathrm{\mho}_2}{\color{black}\check{\mathbf{x}}_p^e}\leq \bar{\mathbf{h}}_{pp,k}^{\mathrm{\mho}_2}{\color{black}\check{\mathbf{x}}_p^e}+\widetilde{\mathbf{h}}_{cp,k}^{\mathrm{\mho}_2}\check{\mathbf{x}}_c-{\color{black}\delta_{p,k}^0}\; , \quad
\forall \|\mathbf{e}_{c,j}\|_2\leq \epsilon_{c,j},\;\forall j\in\mathcal{J}\; .
\end{eqnarray}
Note that
\begin{align}
\|-\tilde{\mathbf{e}}_{p,k}^{\mathrm{\mho}_1}{\color{black}\check{\mathbf{x}}_p^e}\|_2 &\leq\|-\tilde{\mathbf{e}}_{p,k}^{\mathrm{\mho}_1}\|_2\|{\color{black}\check{\mathbf{x}}_p^e}\|_2\\
&=\medmath{\left\|\begin{bmatrix}
\mathcal{R}\{\mathbf{e}_{p,k}^T\}&\mathcal{I}\{\mathbf{e}_{p,k}^T\}
\end{bmatrix}
\begin{bmatrix}
\sin\theta&-\cos\theta\\-\cos\theta&-\sin\theta
\end{bmatrix}\right\|_F\left\|{\color{black}\check{\mathbf{x}}_p^e}\right\|_2}\\
&\medmath{\leq \left\|\begin{bmatrix}
\mathcal{R}\{\mathbf{e}_{p,k}^T\}&\mathcal{I}\{\mathbf{e}_{p,k}^T\}
\end{bmatrix}\right\|_F
\left\|\begin{bmatrix}
\sin\theta&-\cos\theta\\-\cos\theta&-\sin\theta
\end{bmatrix}\right\|_F\left\|{\color{black}\check{\mathbf{x}}_p^e}\right\|_2}\\
&\leq \sqrt{2}\epsilon_{p,k}\|{\color{black}\check{\mathbf{x}}_p^e}\|_2 \; ,\label{normupperbound}
\end{align}
and similarly, we can show  
$
\|-\tilde{\mathbf{e}}_{p,k}^{\mathrm{\mho}_2}{\color{black}\check{\mathbf{x}}_p^e}\|_2 
\leq \sqrt{2}\epsilon_{p,k}\|{\color{black}\check{\mathbf{x}}_p^e}\|_2 \; .
$ 
Thus, if we can guarantee that the following constraints are satisfied, namely 
\begin{align}
\bar{\mathbf{H}}_{pp,k}^{\mathrm{\mho}}{\color{black}\check{\mathbf{x}}_p^e}+\widetilde{\mathbf{H}}_{cp,k}^{\mathrm{\mho}}\check{\mathbf{x}}_c &\geq(\sqrt{2}\epsilon_{p,k}\|{\color{black}\check{\mathbf{x}}_p^e}\|_2+{\color{black}\delta_{p,k}^0})\mathbf{1}_2,\; \forall k\in\mathcal{K} \\
\bar{\mathbf{H}}_{pc,j}^{\mathrm{\mho}}{\color{black}\check{\mathbf{x}}_p^e}+\widetilde{\mathbf{H}}_{cc,j}^{\mathrm{\mho}}\check{\mathbf{x}}_c &\geq(\sqrt{2}\epsilon_{c,j}\|{\color{black}\check{\mathbf{x}}_p^e}\|_2+{\color{black}\delta_{c,j}^0})\mathbf{1}_2,\; \forall j\in\mathcal{J} \; ,
\end{align}
then the constraints in~(\ref{normbound1}) and~(\ref{normbound2}) will be satisfied as well.

Using the above results, we obtain the robust precoder by solving the following optimization problem:
\begin{align}
\minimize_{\check{\mathbf{x}}_c} \;&\|\check{\mathbf{x}}_c\|^2\tag{P2}\\
\text{subject to}\;
&\bar{\mathbf{H}}_{pp,k}^{\mathrm{\mho}}{\color{black}\check{\mathbf{x}}_p^e}+\widetilde{\mathbf{H}}_{cp,k}^{\mathrm{\mho}}\check{\mathbf{x}}_c\geq(\sqrt{2}\epsilon_{p,k}\|{\color{black}\check{\mathbf{x}}_p^e}\|_2+{\color{black} \delta_{p,k}^0})\mathbf{1}_2, \quad \forall k\in\mathcal{K}\tag{C2-1}\label{C2-1}\\
& \bar{\mathbf{H}}_{pc,j}^{\mathrm{\mho}}{\color{black}\check{\mathbf{x}}_p^e}+\widetilde{\mathbf{H}}_{cc,j}^{\mathrm{\mho}}\check{\mathbf{x}}_c\geq(\sqrt{2}\epsilon_{c,j}\|{\color{black}\check{\mathbf{x}}_p^e}\|_2+{\color{black}\delta_{c,j}^0})\mathbf{1}_2\; ,\quad \forall j\in\mathcal{J} \; . \tag{C2-2}\label{C2-2}
\end{align}
As in the case with perfect CSI, the robust SLP design can be found via a quadratic program with linear inequality constraints.

\section{Robust SLP for Stochastic CSI Errors}\label{Stochastic}
The bounded error model above is a very conservative approach, given its goal of ensuring that the SM constraints are met for all possible CSI error realizations. A less conservative approach that allows constraint violations with some acceptably small probability is to assume a statistical CSI error model. As an example, in this section we consider the case where such a model for the PBS CSI error is available due to knowledge of how the channel is quantized. In particular, we assume that the channels $\mathbf{h}_{pp,k}^Q$ and $\mathbf{h}_{pc,j}^Q$ shared by the PBS are element-wise quantized, and we use the approximate additive quantization noise model (AQNM) \cite{bai2013optimization,bai2015energy} to describe their resulting statistics. Other models are possible based on the specific quantization method employed. 

We assume that the channels are Gaussian with zero mean and covariances given by
\begin{align}
\mathbb{R}_{\mathbf{h}_{pp,k}}&\triangleq\mathbb{E}\{\mathbf{h}_{pp,k}^{\text{H}}\mathbf{h}_{pp,k}\}=\beta_p\mathbf{I}_{M_p}\\
\mathbb{R}_{\mathbf{h}_{pc,j}}&\triangleq\mathbb{E}\{\mathbf{h}_{pc,j}^{\text{H}}\mathbf{h}_{pc,j}\}=\beta_c\mathbf{I}_{M_p} \; .
\end{align}
Using AQNM, the quantized CSI from the PBS after rotation is expressed as
\begin{align}
\widetilde{\mathbf{h}}_{pp,k}^Q&=\mathbb{Q}(\widetilde{\mathbf{h}}_{pp,k})\approx\alpha_p\widetilde{\mathbf{h}}_{pp,k}+\widetilde{\mathbf{n}}_{pp,k}^{Q}\label{Qhpp}\\
\widetilde{\mathbf{h}}_{pc,j}^Q&=\mathbb{Q}(\widetilde{\mathbf{h}}_{pc,j})\approx\alpha_c\widetilde{\mathbf{h}}_{pc,j}+\widetilde{\mathbf{n}}_{pc,j}^{Q}\label{Qhpc},
\end{align}
where $\mathbb{Q}(\cdot)$ is a scalar quantization function applied element-wise and separately to the real and imaginary parts of the input. The vectors $\widetilde{\mathbf{n}}_{pp,k}^{Q}\triangleq s_{p,k}^*\mathbf{n}_{pp,k}^Q\in\mathbb{C}^{1\times M_p}$ and $\widetilde{\mathbf{n}}_{pc,j}^{Q}\triangleq s_{c,j}^*\mathbf{n}_{pc,j}^Q\in\mathbb{C}^{1\times M_p}$ denote the zero-mean Gaussian-distributed quantization noise vectors, and both are assumed to be  uncorrelated with $\widetilde{\mathbf{h}}_{pp,k}$ and $\widetilde{\mathbf{h}}_{pc,j}$. 
\begin{table}
\centering
\caption{Distortion Factors for Different Quantization Bit Resolutions \cite{max1960quantizing}}
\begin{tabular}{|c|c|c|c|c|c|}
\hline
$b$ &1  &2  &3  &4&5  \\
\hline
$\rho$& 0.3634&0.1175&0.03454&0.009497&0.002499\\
\hline
\end{tabular}
\label{table:1}
\end{table}
The gains $\alpha_k=1-\rho_k$ for $k\in\{p,c\}$ are functions of the following distortion factors \cite{fletcher2007robust}:
 \begin{equation}
\rho_p=\frac{\mathbb{E}\{\|\mathbf{h}_{pp,k}-\mathbf{h}_{pp,k}^Q\|^2\}}{\mathbb{E}\{\|\mathbf{h}_{pp,k}\|^2\}},\; \rho_c=\frac{\mathbb{E}\{\|\mathbf{h}_{pc,j}-\mathbf{h}_{pc,j}^Q\|^2\}}{\mathbb{E}\{\|\mathbf{h}_{pc,j}\|^2\}} \; .
\end{equation}
The value of $\rho$ is given in Table \ref{table:1} for different bit resolutions $b$ assuming an optimal non-uniform Lloyd-Max quantizer \cite{max1960quantizing}. 
The phase rotation does not alter the covariance matrices of the quantization noise, which are given by \cite{bai2013optimization}
\begin{align}
\mathbb{R}_{\widetilde{\mathbf{n}}_{pp,k}^{Q}}&
=\alpha_p\rho_p\text{diag}\{\mathbb{R}_{\mathbf{h}_{pp,k}}\}=\alpha_p\rho_p\beta_p\mathbf{I}_{M_p}\label{nppVariance},\\
\mathbb{R}_{\widetilde{\mathbf{n}}_{pc,j}^{Q}}&
=\alpha_c\rho_c\text{diag}\{\mathbb{R}_{\mathbf{h}_{pc,j}}\}=\alpha_c\rho_c\beta_c\mathbf{I}_{M_p}\label{npcVariance}.
\end{align}

Based on Eq.~(\ref{Qhpp}) and Eq.~(\ref{Qhpc}), we can derive
\begin{eqnarray}
\widetilde{\mathbf{h}}_{pp,k}=\frac{\widetilde{\mathbf{h}}_{pp,k}^Q-\widetilde{\mathbf{n}}_{pp,k}^Q}{\alpha_p}=\bar{\alpha_p}\widetilde{\mathbf{h}}_{pp,k}^Q-\bar{\alpha_p}\widetilde{\mathbf{n}}_{pp,k}^Q\; , \\
\widetilde{\mathbf{h}}_{pc,j}=\frac{\widetilde{\mathbf{h}}_{pc,j}^Q-\widetilde{\mathbf{n}}_{pc,j}^Q}{\alpha_c}=\bar{\alpha_c}\widetilde{\mathbf{h}}_{pc,j}^Q-\bar{\alpha_c}\widetilde{\mathbf{n}}_{pc,j}^Q\; ,
\end{eqnarray}
where $\bar{\alpha_p}=\frac{1}{\alpha_p}$ and $\bar{\alpha_c}=\frac{1}{\alpha_c}$. Therefore, 
\begin{eqnarray}
\widetilde{\mathbf{H}}_{pp,k}^{\mathrm{\mho}}=\mathrm{\mho}(\bar{\alpha_p}\widetilde{\mathbf{h}}_{pp,k}^Q-\bar{\alpha_p}\widetilde{\mathbf{n}}_{pp,k}^Q)=\bar{\alpha_p}(\widetilde{\mathbf{H}}_{pp,k}^{Q,\mathrm{\mho}}-\widetilde{\mathbf{N}}_{pp,k}^{Q,\mathrm{\mho}})\label{tqpp}\\
\widetilde{\mathbf{H}}_{pc,j}^{\mathrm{\mho}}=\mathrm{\mho}(\bar{\alpha_c}\widetilde{\mathbf{h}}_{pc,j}^Q-\bar{\alpha_c}\widetilde{\mathbf{n}}_{pc,j}^Q)=\bar{\alpha_c}(\widetilde{\mathbf{H}}_{pc,j}^{Q,\mathrm{\mho}}-\widetilde{\mathbf{N}}_{pc,j}^{Q,\mathrm{\mho}})\label{tqpc},
\end{eqnarray}
where $\widetilde{\mathbf{H}}_{pp,k}^{Q,\mathrm{\mho}}\triangleq\mathrm{\mho}(\widetilde{\mathbf{h}}_{pp,k}^Q)$, $\widetilde{\mathbf{N}}_{pp,k}^{Q,\mathrm{\mho}}\triangleq\mathrm{\mho}(\widetilde{\mathbf{n}}_{pp,k}^Q)$, $\widetilde{\mathbf{H}}_{pc,j}^{Q,\mathrm{\mho}}\triangleq\mathrm{\mho}(\widetilde{\mathbf{h}}_{pc,j}^Q)$, and $\widetilde{\mathbf{N}}_{pc,j}^{Q,\mathrm{\mho}}\triangleq\mathrm{\mho}(\widetilde{\mathbf{n}}_{pc,j}^Q)$.
Substituting Eq.~(\ref{tqpp}) and Eq.~(\ref{tqpc}) in (\ref{C1}), we have
\begin{eqnarray}
\bar{\alpha_p}(\widetilde{\mathbf{H}}_{pp,k}^{Q,\mathrm{\mho}}-\widetilde{\mathbf{N}}_{pp,k}^{Q,\mathrm{\mho}})\check{\mathbf{x}}_p+\widetilde{\mathbf{H}}_{cp,k}^{\mathrm{\mho}}\check{\mathbf{x}}_c\geq{\color{black}\delta_{p,k}^0}\mathbf{1}_2, \forall k\in\mathcal{K}\label{stoCons1}\\
\widetilde{\mathbf{H}}_{cc,j}^{\mathrm{\mho}}\check{\mathbf{x}}_c+\bar{\alpha_c}(\widetilde{\mathbf{H}}_{pc,j}^{Q,\mathrm{\mho}}-\widetilde{\mathbf{N}}_{pc,j}^{Q,\mathrm{\mho}})\check{\mathbf{x}}_p\geq{\color{black}\delta_{c,j}^0}\mathbf{1}_2, \forall j\in\mathcal{J}.\label{stoCons2}
\end{eqnarray}

{\color{black}
\subsection{Primary System} 
As a special case to fix the details, we assume that the PBS employs ZF precoding to cancel the interference among the PUs. Thus, the transmit symbol at the PBS can be expressed as
\begin{equation}
\mathbf{x}_p=f_p\mathbf{H}_{pp}^H(\mathbf{H}_{pp}\mathbf{H}_{pp}^H)^{-1}\mathbf{s}_p\; ,
\end{equation}
where $f_p=\sqrt{\frac{P_p}{\text{trace}\{(\mathbf{H}_{pp}\mathbf{H}_{pp}^H)^{-1}\}}}$ is the scaling factor to satisfy the PBS power budget. Then
\begin{align}
(\widetilde{\mathbf{H}}_{pp,k}^{Q,\mathrm{\mho}}-\widetilde{\mathbf{N}}_{pp,k}^{Q,\mathrm{\mho}})\check{\mathbf{x}}_p&=\alpha_p\widetilde{\mathbf{H}}_{pp,k}^{\mathrm{\mho}}\check{\mathbf{x}}_p\\
&=\alpha_p\begin{bmatrix}\mathcal{R}\{\tilde{\mathbf{h}}_{pp,k}\mathbf{x}_p\}\sin \theta-\mathcal{I}\{\tilde{\mathbf{h}}_{pp,k}\mathbf{x}_p\}\cos\theta\\\mathcal{R}\{\tilde{\mathbf{h}}_{pp,k}\mathbf{x}_p\}\sin \theta+\mathcal{I}\{\tilde{\mathbf{h}}_{pp,k}\mathbf{x}_p\}\cos\theta\end{bmatrix}\\
&=\alpha_pf_p\sin\theta\mathbf{1}_2
\label{qpk}
\end{align}
due to $\tilde{\mathbf{h}}_{pp,k}\mathbf{x}_p=s_{p,k}^*f_ps_{p,k}=f_p$, which is not surprising since, even with imperfect CSI, the CBS can assume the ZF precoding at the PBS is successful in delivering the desired symbols to the users. The exact value of the scaling factor $f_p$ depends on the true channel $\mathbf{H}_{pp}$, but the CBS can employ an estimate based on its quantized approximation:  
\begin{equation}
f_p^Q=\sqrt{\frac{P_p}{\text{trace}\{(\mathbf{H}_{pp}^Q(\mathbf{H}_{pp}^Q)^H)^{-1}\}}} \; ,
\end{equation}
where $\mathbf{H}_{pp}^Q=\begin{bmatrix} (\mathbf{h}_{pp,1}^Q)^{\text{T}}&\cdots&(\mathbf{h}_{pp,N_p}^Q)
^{\text{T}}\end{bmatrix}^{\text{T}}$, and $P_p$ is assumed to be known. Using a similar argument, we can obtain the following deterministic form of the constraint in Eq.~(\ref{stoCons1}) as follows:
\begin{equation}
\widetilde{\mathbf{H}}_{cp,k}^{\mathrm{\mho}}\check{\mathbf{x}}_c\geq(\delta_{p,k}^0-f_p^Q\sin\theta)\mathbf{1}_2\; ,\label{stoP1}
\end{equation}

\subsection{Cognitive System}
For the cognitive system, the constraint~(\ref{stoCons2}) above is expressed in terms of the unknown random quantization noise, and thus cannot be directly enforced. Instead, we choose to pose the problem such that the constraint is achieved with a certain probability. In particular, considering that $\check{\mathbf{x}}_p$ relies on $\mathbf{H}_{pp,k}$ and thus is also uncertain, we rewrite~(\ref{stoCons2}) as follows:
\begin{equation}
\mathbb{P}\{\alpha_c(\widetilde{\mathbf{H}}_{cc,j}^{\mathrm{\mho}}\check{\mathbf{x}}_c-\delta_{c,j}^0\mathbf{1}_2)\geq(\widetilde{\mathbf{N}}_{pc,j}^{Q,\mathrm{\mho}}-\widetilde{\mathbf{H}}_{pc,j}^{Q,\mathrm{\mho}})\check{\mathbf{x}}_p\}\geq v_c\label{pcc}\; ,
\end{equation}
where $\mathbb{P}\{A\}$ denotes the probability of event $A$, and $v_c\in(0.5,1]$ represent the probability threshold. In the following, we find expressions for the probabilities in~(\ref{pcc}).

First, we get
\begin{eqnarray}
\mathbb{E}\{\widetilde{\mathbf{N}}_{pc,j}^{Q,\mathrm{\mho}}\} &=& \mathbf{0}_{2\times 2M_p}\\
\mathbb{E}\{\widetilde{\mathbf{N}}_{pc,j}^{Q,\mathrm{\mho}}(\widetilde{\mathbf{N}}_{pc,j}^{Q,\mathrm{\mho}})^{\text{H}}\} &= & M_p\alpha_c\rho_c\beta_c
\begin{bmatrix}
1&-\cos2\theta\\
-\cos2\theta&1
\end{bmatrix}\\
\mathbb{E}\{\widetilde{\mathbf{H}}_{pc,j}^{Q,\mathrm{\mho}}(\widetilde{\mathbf{H}}_{pc,j}^{Q,\mathrm{\mho}})^H\}&=&M_p\alpha_c\beta_c\begin{bmatrix}
1&-\cos2\theta\\
-\cos2\theta&1
\end{bmatrix}\; ,
\end{eqnarray}
and we define
\begin{equation}
\mathbf{q}_{c,j}\triangleq(\widetilde{\mathbf{N}}_{pc,j}^{Q,\mathrm{\mho}}-\widetilde{\mathbf{H}}_{pc,j}^{Q,\mathrm{\mho}})\check{\mathbf{x}}_p\triangleq\begin{bmatrix}q_{c,j}^1\\q_{c,j}^2\end{bmatrix}\; .\label{qcj}
\end{equation}
We can show that $\mathbf{q}_{c,j}$ is a bivariate correlated Gaussian random variable with mean
\begin{equation}
\mathbb{E}\{\mathbf{q}_{c,j}\}=\mathbf{0}_{2\times 1}
\end{equation}
and covariance
\begin{align}
\mathbb{R}_{\mathbf{q}_{c,j}}&=\mathbb{E}\{(\widetilde{\mathbf{N}}_{pc,j}^{Q,\mathrm{\mho}}-\widetilde{\mathbf{H}}_{pc,j}^{Q,\mathrm{\mho}})\check{\mathbf{x}}_p\check{\mathbf{x}}_p^H(\widetilde{\mathbf{N}}_{pc,j}^{Q,\mathrm{\mho}}-\widetilde{\mathbf{H}}_{pc,j}^{Q,\mathrm{\mho}})^H\}\\
&=\frac{P_p}{2M_p}\left\{\mathbb{E}\{\widetilde{\mathbf{N}}_{pc,j}^{Q,\mathrm{\mho}}(\widetilde{\mathbf{N}}_{pc,j}^{Q,\mathrm{\mho}})^H\}+\mathbb{E}\{\widetilde{\mathbf{H}}_{pc,j}^{Q,\mathrm{\mho}}(\widetilde{\mathbf{H}}_{pc,j}^{Q,\mathrm{\mho}})^H\}\right\}\\
&=\frac{P_p\beta_c\alpha_c(2-\alpha_c)}{2}
\begin{bmatrix}
1&-\cos2\theta\\
-\cos2\theta&1
\end{bmatrix}\label{Rqc}\; .
\end{align}
Furthermore, we define
\begin{equation}
\mathbf{w}_{c,j}(\check{\mathbf{x}}_c)\triangleq \alpha_c\widetilde{\mathbf{H}}_{cc,j}^{\mathrm{\mho}}\check{\mathbf{x}}_c-\alpha_c{\color{black}\delta_{c,j}^0}\mathbf{1}_2\triangleq\begin{bmatrix}w_{c,j}^1\\w_{c,j}^2\end{bmatrix}
\end{equation}
which is affine in $\check{\mathbf{x}}_c$. Using the new notation, the chance constraint (\ref{pcc}) can be rewritten as
\begin{equation}
\mathbb{P}\{\mathbf{w}_{c,j}(\check{\mathbf{x}}_c)\geq\mathbf{q}_{c,j}\}\geq v_c\label{cconstraint}\; .
\end{equation}
For ease of notation, we define $\bar{\mathbf{w}}_{c,j}(\check{\mathbf{x}}_c)\triangleq\mathbb{R}_{\mathbf{q}_{c,j}}^{-\frac{1}{2}}\mathbf{w}_{c,j}(\check{\mathbf{x}}_c)$ and $\bar{\mathbf{q}}_{c,j}\triangleq\mathbb{R}_{\mathbf{q}_{c,j}}^{-\frac{1}{2}}\mathbf{q}_{c,j}$, and we obtain the following lemma.
\begin{lemma}
\label{coglemma}
$\mathbb{P}\{\bar{\mathbf{w}}_{c,j}(\check{\mathbf{x}}_c)\geq\bar{\mathbf{q}}_{c,j}\}\geq v_c$ can be approximated by the inequality 
\begin{equation}
\label{stoP2}
   \widetilde{\mathbf{H}}_{cc,j}^{\mathrm{\mho}}\check{\mathbf{x}}_c\geq \bar{\alpha_c}\eta_c\mathbb{R}_{\mathbf{q}_{c,j}}^{\frac{1}{2}}\mathbf{1}_2+{\color{black}\delta_{c,j}^0}\mathbf{1}_2,
\end{equation}
where $\eta_c=\sqrt{2}\erf^{-1}\left(2\sqrt{v_c}-1\right)$ is a preset constant.
\end{lemma}
\begin{proof}
See Appendix \ref{lemma1proof}. 
\end{proof}
With this lemma, knowledge of the precise value for $\mathbf{x}_p$ is not necessary in the design of the precoder at the CBS, which is important under the assumption of a finite capacity channel for information sharing.

\subsection{Optimization Problem for Stochastic CSI Error Model}
We can now formulate the robust SLP design with probabilistic constraints by replacing ~(\ref{stoCons1}) and~(\ref{stoCons2}) 
with~(\ref{stoP1}) and~(\ref{stoP1}), as follows:
\begin{align}
\minimize_{\check{\mathbf{x}}_c} \quad&\|\check{\mathbf{x}}_c\|^2\tag{P3}\\
\text{subject to}\quad 
&\widetilde{\mathbf{H}}_{cp,k}^{\mathrm{\mho}}\check{\mathbf{x}}_c\geq(\delta_{p,k}^0-f_p^Q\sin\theta)\mathbf{1}_2\;, \forall k\in\mathcal{K}\tag{C3-1}\; ,\\
&   \widetilde{\mathbf{H}}_{cc,j}^{\mathrm{\mho}}\check{\mathbf{x}}_c\geq \bar{\alpha_c}\eta_c\mathbb{R}_{\mathbf{q}_{c,j}}^{\frac{1}{2}}\mathbf{1}_2+{\color{black}\delta_{c,j}^0}\mathbf{1}_2\;, \forall j\in\mathcal{J}\tag{C3-2} \; .
\end{align}
As with the previous problem studied above, the result is a quadratic program with linear inequality constraints which is robust to imperfect CSI shared from the PBS.
}

\section{Numerical Results}\label{numresult}
In this section, we assess the performance of our proposed power-minimizing SLP (PMSLP) approaches. Monte-Carlo simulations are conducted over 1000 independent channel realizations, each employing a block of $T=100$ symbols. The channels $\mathbf{H}_{pp},\ \mathbf{H}_{cp},\ \mathbf{H}_{pc}$ and  $\mathbf{H}_{cc}$ are composed of i.i.d. Gaussian random variables with zero mean and unit variance. The complex Gaussian noise is assumed to have the same power ($\sigma_p=\sigma_c=1$) for all PUs and CUs. The PBS transmission power is set at $P_p=10$ dBW. {\color{black} We employ the same threshold for all users within a given network, i.e., $\delta_{p,1}^0=\cdots=\delta_{p,N_p}^0=\delta_p^0$ for the PUs and $\delta_{c,1}^0=\cdots=\delta_{c,N_c}^0=\delta_c^0$ for the CUs, ensuring the same worst-case SER for the users in each network.}

Since for SLP we work with finite alphabet constellations, we will analyze the block transmission performance of the system using the throughput $\tau$ as calculated in \cite{salem2021error}:
\begin{equation}\label{throughput}
\tau=(1-P_B)\times c\times T\times N , 
\end{equation}
where $P_B$ is the block error rate (BLER),  $c=\log_2D$ is the number of bits per modulation symbol, $T$ is the block length and $N$ is the number of receivers. In each block for each user, there are $C=c\times T$ data message bits transmitted from the BS. For PSK modulation, assuming a binomial distribution of errors in each block, the probability of more than $q$ errors occurring in one block of $C$ bits is expressed as 
\begin{equation}
        P_e(q,C)=1-\sum_{i=0}^{q}\binom{C}{i}P_b^i(1-P_b)^{C-i}
\end{equation}
where $P_b$ is the BER. If the receiver detects errors without correction, a block is received correctly only if all $C$ bits in the block are received correctly, and thus the BLER is $P_B=P_e(0,C)$. On the other hand, if the receiver is capable of correcting up to $Q$ errors in each block, then the BLER is given by $P_B=P_e(Q,C)$ \cite{eaves1977probability}. 

\begin{figure}[!t]
\centering
\includegraphics[width=3in]{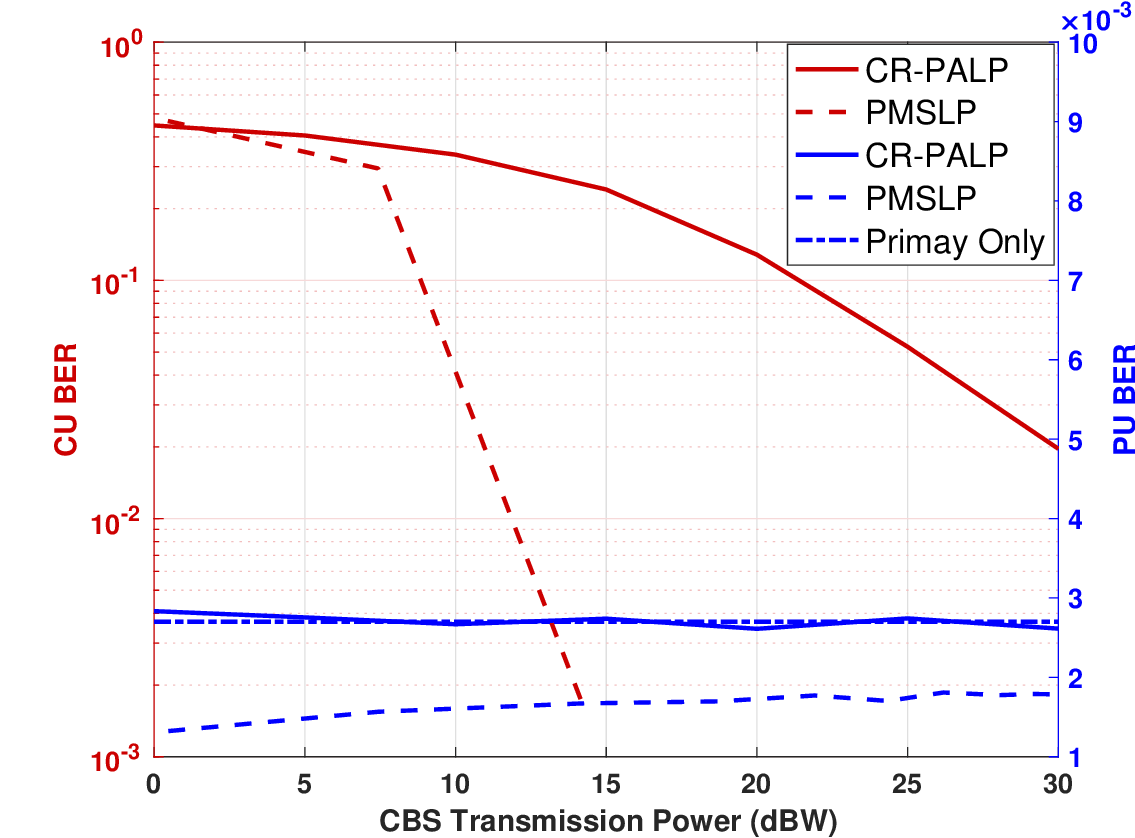}
\caption{BER of CU (left) or PU (right) vs. CBS transmit power where $P_p=10\ \text{dBW}$, QPSK modulation.}
\label{PerfectCSI}
\end{figure}

We begin in Fig.~ \ref{PerfectCSI} assuming perfect CSI, plotting the average BER of the users versus the CBS transmission power, and comparing PMSLP assuming a minimum safety margin $\delta_p^0=1.9$ with the performance of the CR-PALP algorithm described in Section \ref{PALPexpression}. 
Both the PBS and CBS are assumed to have $M_p=M_c=8$ antennas and the number of PUs and CUs are both set at $N_p=N_c=4$. 
With these settings, even when the CBS increases its transmit power to better serve the CUs, it can still avoid any negative impact on the PUs such that the BER of PUs is not greater than that in the primary-only case as shown in Fig.~ \ref{PerfectCSI}. Moreover, the BER of the PUs remains nearly unchanged for both types of precoders, although the PUs actually enjoy some benefit with PMSLP since it exploits CI from the CBS signals which can further improve the SM for the PUs. Meanwhile, PMSLP provides a much lower BER for the CUs; the CBS can save more than $15$dBW of power to achieve an uncoded BER of $10^{-2}$, compared to the CR-PALP method which allows only the CUs to benefit from the inter-user CI in the cognitive system, but in general will not entirely eliminate the destructive interference from the PBS. On the contrary, our proposed PMSLP precoder can take advantage of CI not only from the CU MUI but also what is available (although not optimized) from the primary system. Moreover, the PUs can also benefit from the CI produced by the cognitive system. CR-PALP requires that the CBS act as a relay transmitting not only the CUs' but also the PUs' signals, which is unnecessary in our PMSLP design. 

\begin{figure}[!t]
\centering
\includegraphics[width=2.7in]{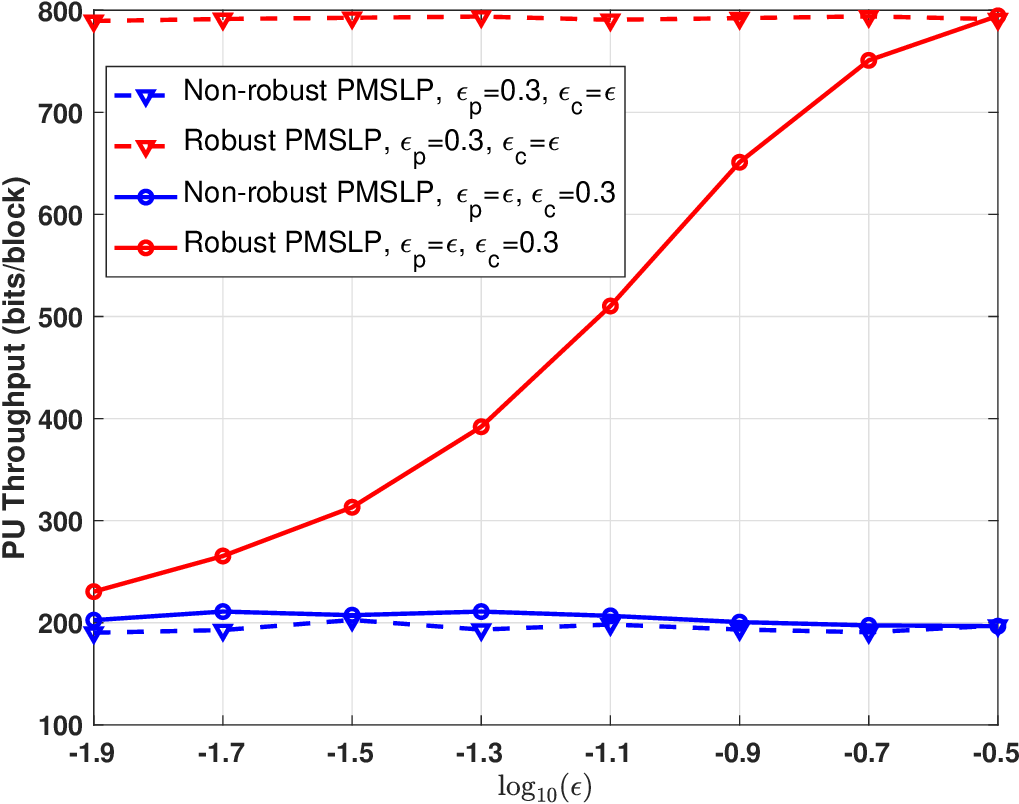}
\caption{{\color{black}Throughput} of PU vs. error norm bound where $P_p=10\ \text{dBW}$, $\delta_p^0=\delta_c^0=1.5$, QPSK modulation.}
\label{BoundEpsilonPrimaryThroughput}
\end{figure}
\begin{figure}[!t]
\centering
\includegraphics[width=2.7in]{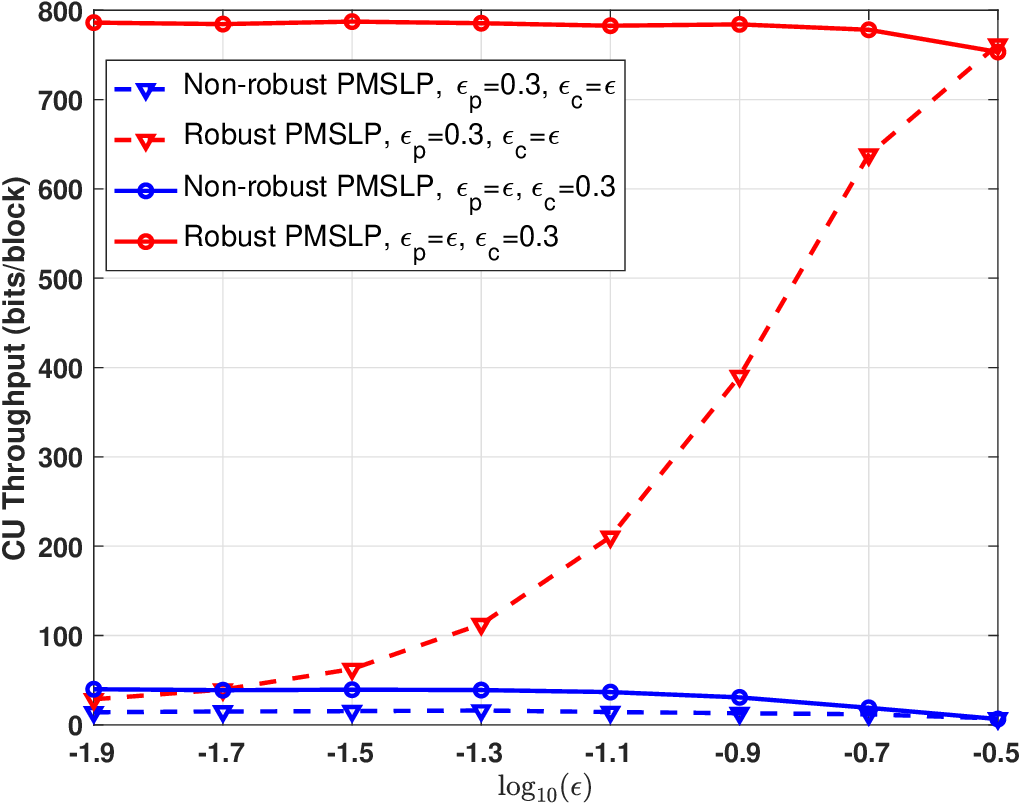}
\caption{{\color{black}Throughput} of CU vs. error norm bound where $P_p=10\ \text{dBW}$, $\delta_p^0=\delta_c^0=1.5$, QPSK modulation.}
\label{BoundEpsilonCognitiveThroughput}
\end{figure}
\begin{figure}[!t]
\centering
\includegraphics[width=2.7in]{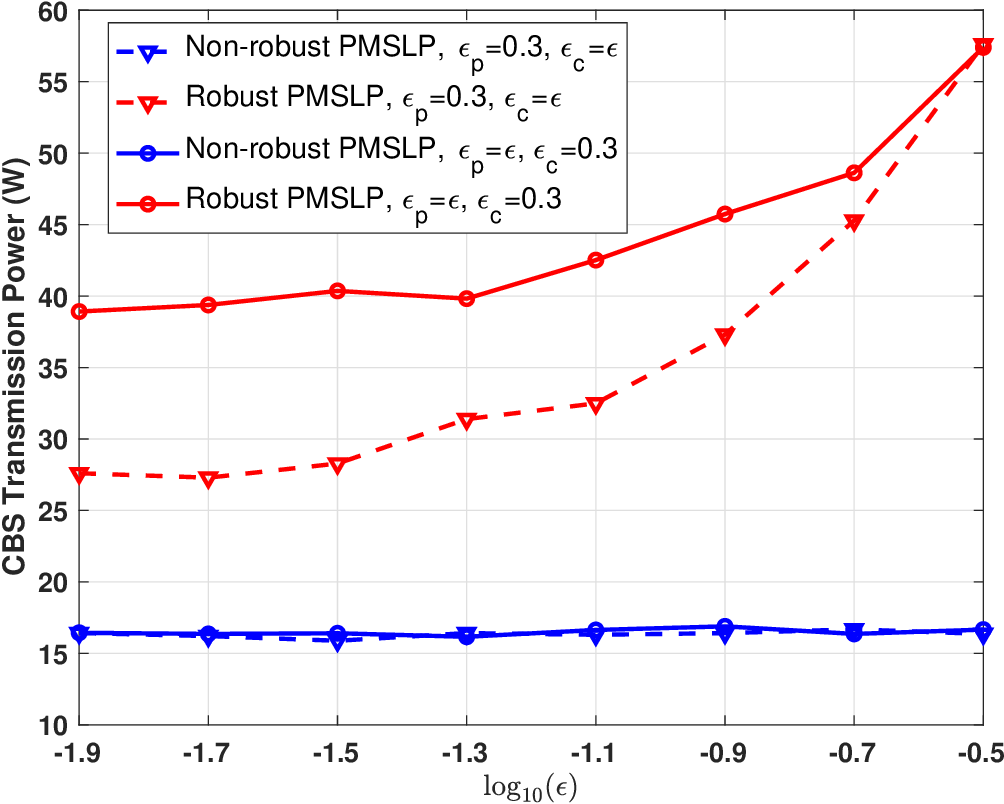}
\caption{Transmit power vs. error norm bound where $P_p=10\ \text{dBW}$, $\delta_p^0=\delta_c^0=1.5$, QPSK modulation.}
\label{BoundEpsilonPower}
\end{figure}
Next we consider the norm-bounded CSI error model discussed in Section \ref{NormBound}. With the SM threshold for the PUs and CUs set to 1.5, we plot the throughput of the PUs and CUs as the norm of one error ($\epsilon_{p,k}$ or $\epsilon_{c,j}$) changes as $\log_{10}\epsilon$ while the norm of the other is fixed to 0.3 {\color{black} {\cite{masouros2015exploiting}}}. For simplicity, we assume that the CSI error bounds are the same for all users: $\epsilon_{p,k}=\epsilon_{p}$ and $\epsilon_{c,j}=\epsilon_{c}$. Fig.~\ref{BoundEpsilonPrimaryThroughput} and Fig.~\ref{BoundEpsilonCognitiveThroughput} respectively show the {\color{black}throughput} for the PUs and CUs as a function of the error bound, and demonstrate that the proposed robust precoder can mitigate the CSI uncertainty and provide a much {\color{black}higher throughput} compared to the non-robust precoder. With the robustness introduced, the {\color{black}throughput} of the PUs and CUs actually {\color{black}increases} as the norm of the corresponding error increases. This can be explained by examining~(\ref{C2-1}) and~(\ref{C2-2}), where we see that a larger error bound creates a larger effective SM in the constraint, which provides the robustness necessary to account for the imperfect channel {\color{black}and also imperfect knowledge of $\mathbf{x}_p$ and $f_p$}. There is however a price to be paid for this robustness, as clearly seen in Fig.~\ref{BoundEpsilonPower}, which shows that the robust schemes require the CBS to operate with significantly more power, especially as the error bound increases. It is clear from these results that the worst-case approach based on the norm-bounded CSI error leads to a conservative design.
\begin{figure}[!t]
\centering
\includegraphics[width=2.7in]{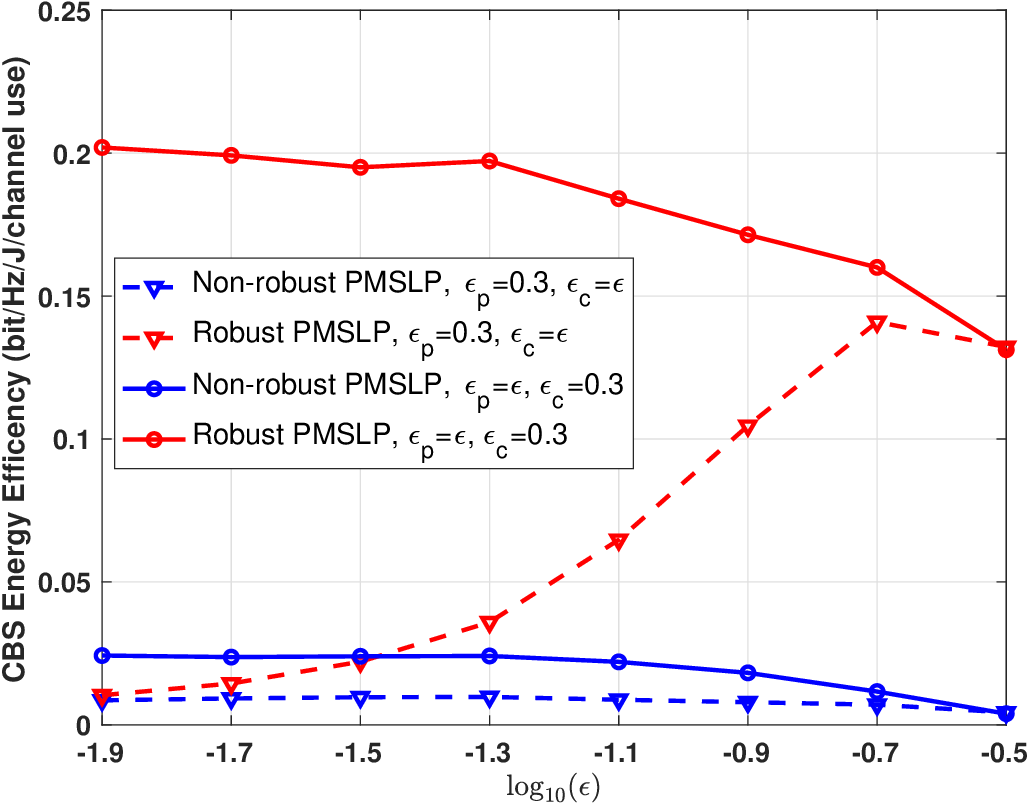}
\caption{Energy Efficiency vs. Norm of the Error where $P_p=10\ \text{dBW}$, $\delta_p^0=\delta_c^0=1.5$, QPSK.}
\label{BoundEpsilonEE}
\end{figure}

In order to quantify the power-performance trade-off between the robust and non-robust designs, in Fig.~\ref{BoundEpsilonEE} we plot the energy efficiency (EE) of the approaches, defined as the ratio between the throughput calculated from Eq.~(\ref{throughput}) and the transmit power per channel:
\begin{equation}
    \text{EE}=\frac{\tau}{T\times \|\check{\mathbf{x}}_c\|^2}.
\end{equation}
We see that despite the increase in transmit power, the proposed robust SLP algorithm achieves a significantly higher energy efficiency. When the uncertainty $\epsilon_c$ in $\mathbf{H}_{pc}$ is fixed, the energy efficiency at the CBS decreases with greater uncertainty in $\mathbf{H}_{pp}$ since the CBS needs to consume more power to meet the SM constraint at the PUs. 



\begin{figure}[!t]
\centering
\includegraphics[width=2.7in]{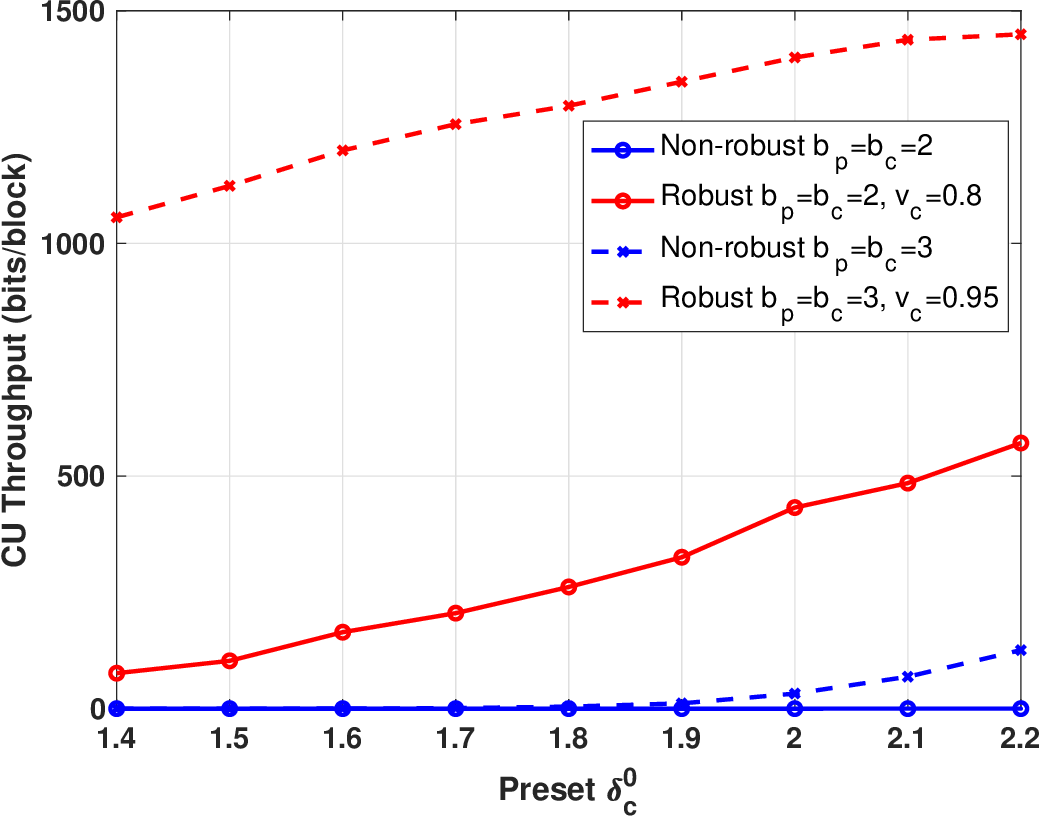}
\caption{{\color{black}Throughput} of CUs vs. preset SM at the CUs where $P_p=10\ \text{dBW}$, $\delta_p^0=1.5$, {\color{black}$Q=1$}, QPSK modulation.}
\label{StochasticSMvsThroughputCU}
\end{figure}

\begin{figure}[!t]
\centering
\includegraphics[width=2.7in]{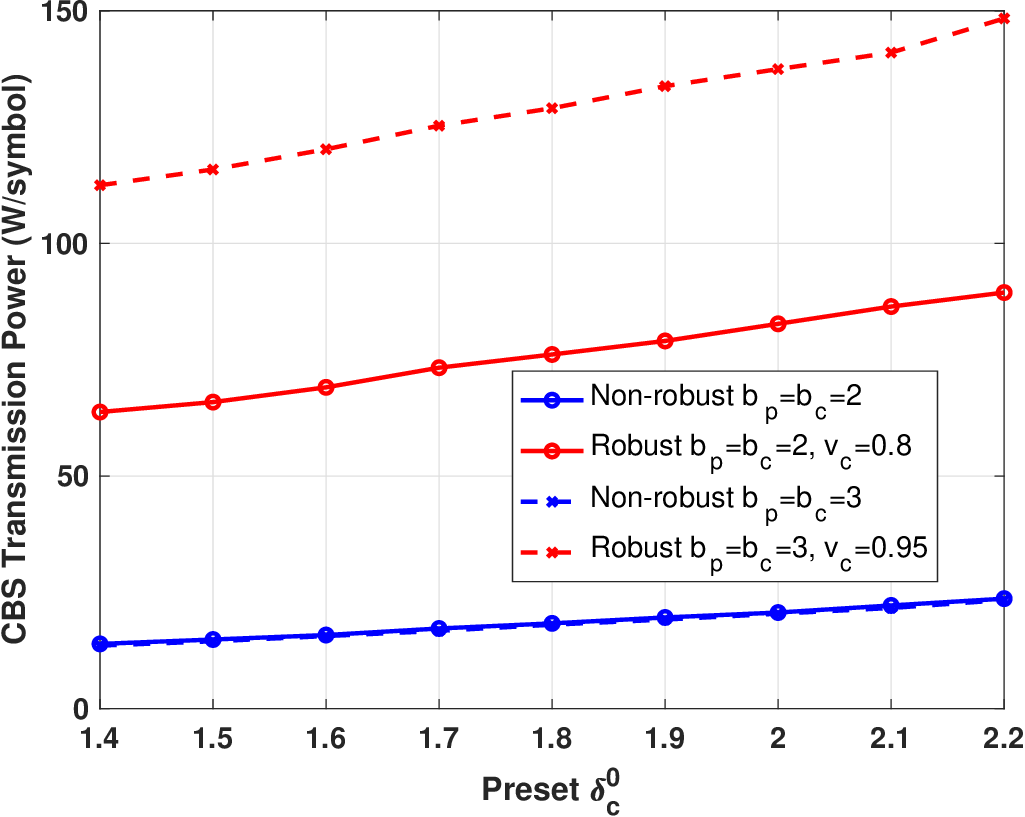}
\caption{Transmit Power at CBS vs. preset SM at the CUs where $P_p=10\ \text{dBW}$, $\delta_p^0=1.5$, {\color{black}$Q=1$}, QPSK modulation.}
\label{StochasticSMvsPower}
\end{figure}

\begin{figure}[!t]
\centering
\includegraphics[width=2.7in]{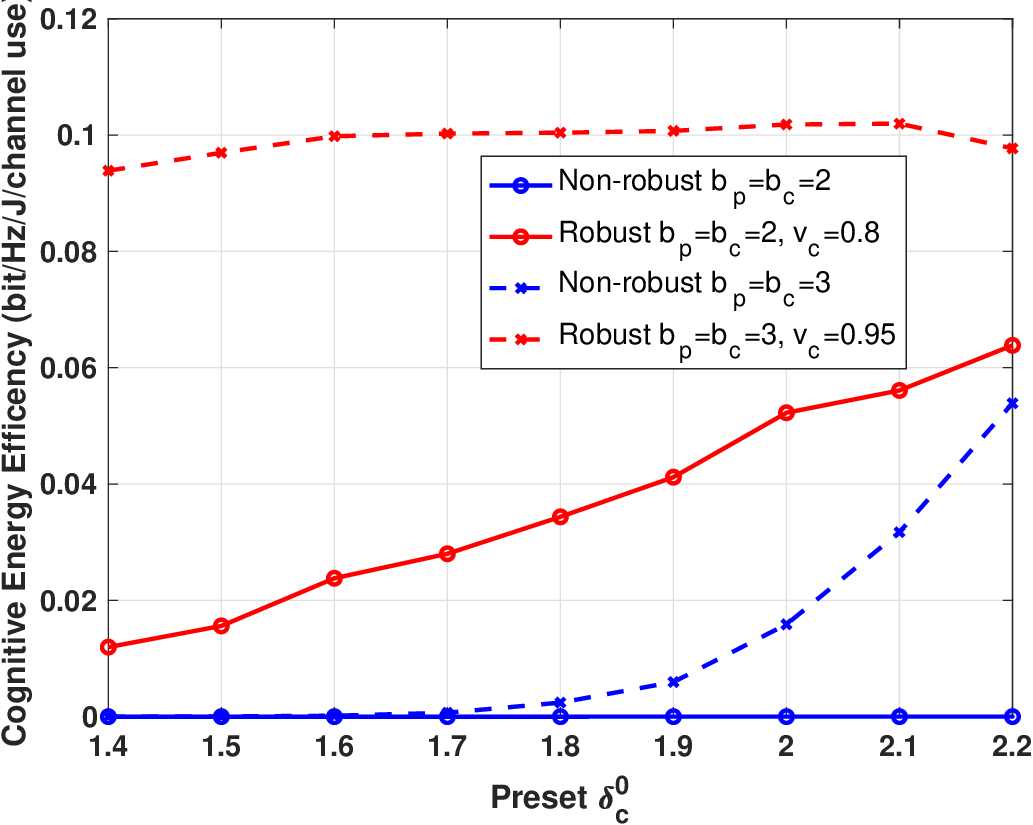}
\caption{EE at CBS vs. preset SM at the CUs where $P_p=10\ \text{dBW}$, $\delta_p^0=1.5$, {\color{black}$Q=1$}, QPSK modulation.}
\label{StochasticSMvsEE}
\end{figure}

The remaining examples use the probabilistic SM constraints discussed in Section~\ref{Stochastic} based on the AQNM approximation, although the actual quantized CSI is generated using a non-uniform Lloyd Max quantizer \cite{max1960quantizing,lloyd1982least}.
{\color{black} In this case, the PBS and CBS are assumed to have $M_p=M_c=16$ antennas and the number of PUs and CUs are both set at $N_p=N_c=8$. The receiver is capable of correcting $Q=1$ bit error in each block \cite{salem2021error}. The probability $v$ was set with the value used in \cite{haqiqatnejad2020robust}.}  
Fig.~\ref{StochasticSMvsThroughputCU} shows the {\color{black}throughput} of the CUs as a function of the preset SM threshold at the CUs, assuming either $b=2$ or $b=3$ quantization bits per channel coefficient and different probability constraints. We see that the CUs reap benefits from the robust SLP design, achieving significantly {\color{black}higher throughput}. Again illustrating the trade-off of robustness with increased power, we see in Fig.~\ref{StochasticSMvsPower} that as the preset $\delta_c^0$ increases, the CBS in the robust SLP approach requires more power to meet the SM constraint than the non-robust SLP. In order to fairly compare different SLP methods, we plot the EE at the CBS in Fig.~\ref{StochasticSMvsEE}. It is clear that the greater the preset SM, or the higher the quantization resolution, the higher the EE. {\color{black} For the case of $b=2$, the EE of the non-robust SLP is nearly 0, but the robust SLP approach performs particularly well even with very low-resolution CSI}.


\begin{figure}[!t]
\centering
\includegraphics[width=2.7in]{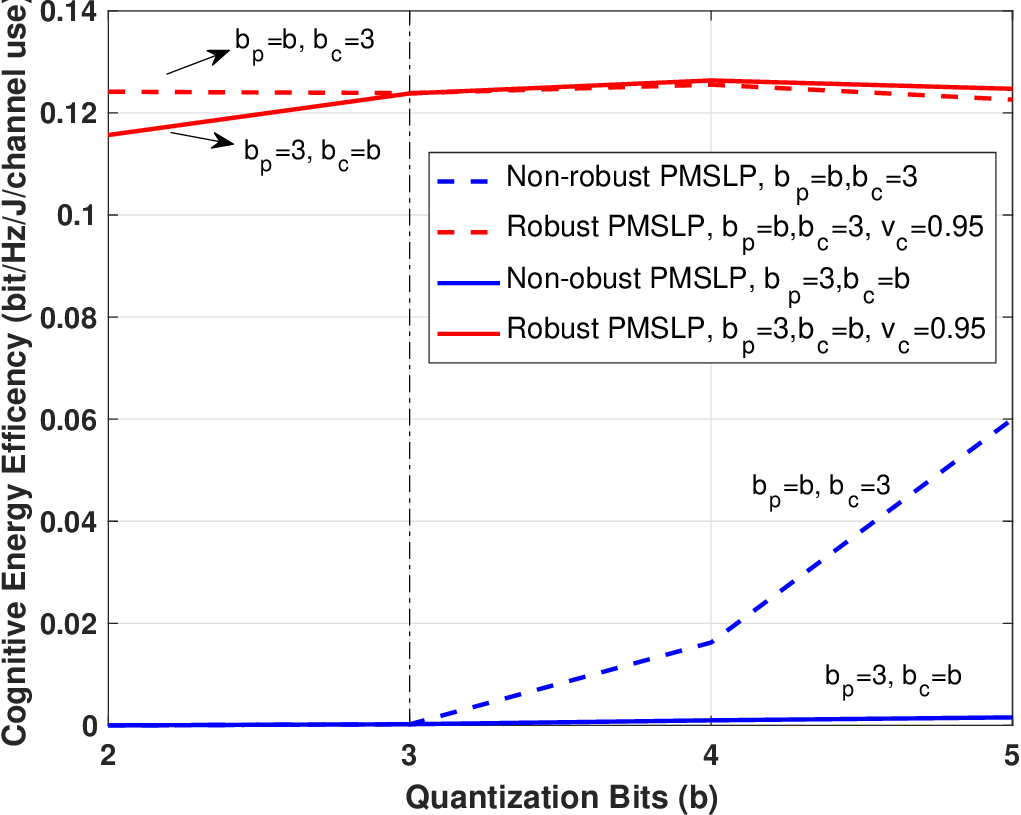}
\caption{Energy Efficiency at the CBS vs. quantization resolution where $P_p=10\ \text{dBW}$, $\delta_p^0=\delta_c^0=1.5$, {\color{black} $Q=1$,} QPSK modulation.}
\label{StochasticBitAllocation}
\end{figure}

In the final example, we study the allocation of the quantization bits on the system performance  \cite{yang2018limited}. In particular, in Fig.~\ref{StochasticBitAllocation} we plot the energy efficiency at the CBS when the direct $\mathbf{h}_{pp,k}$ and interference channels $\mathbf{h}_{pc,j}$ are quantized with different resolutions. {\color{black} For the non-robust SLP schemes (blue curves), the CBS achieves higher EE in cases where $b_p>b_c$, indicating that for a fixed number of quantization bits, it is more energy efficient for the cognitive system to receive a more accurate representation of the direct channel than the interference channel.} However, the robust-SLP schemes are less sensitive to the allocation of the quantization bits, and show roughly the same performance regardless of which channel is more accurately represented.

\section{Conclusion}\label{conclusion}
In this paper, we have designed non-robust and robust SLP schemes for overlay CR systems with the goal of minimizing the transmission power and simultaneously ensuring the QoS of all users. Unlike traditional CR precoding techniques, we set the SM threshold in the interference constraints instead of using SINR or BER metrics in order to fully exploit constructive interference as much as possible. First, under the assumption of perfect CSI, we propose an SLP algorithm that performs significantly better than a prior CR-based SLP approach modified to address our overlay problem. In the proposed algorithm, not only the CUs but also the PUs benefit from the constructive interference. Then, using two different CSI error models, we derive two robust SLP methods, one based on a max-min optimization of the worst-case CSI error and the other on a probability-constrained
problem using AQNM to approximate the impact of the CSI quantization. All of the proposed optimization problems result in a quadratic objective function with linear inequality constraints that can be efficiently solved. Our numerical results demonstrate that our robust SLP schemes can deal with various types of CSI error and still maintain a high energy efficiency. A key observation from our results is that, by enabling the PUs to exploit constructive interference as well as the CUs, the presence of the cognitive network can actually {\em improve} the PU network performance rather than degrade it.


\appendix
\subsection{Proof of Lemma \ref{coglemma}}\label{lemma1proof}

From \cite{haqiqatnejad2020robust} we note that eliminating the (possible) correlation between the entries of $\mathbf{q}_{c,j}$ by applying a whitening transform can ease the difficulty of finding the desired approximation. In particular, we will apply the whitening matrix from \cite{kessy2018optimalwhitening} which is optimal in terms of mean-squared error, i.e.,
{\color{black}
\begin{equation}
\mathbb{R}_{\mathbf{q}_{c,j}}^{-\frac{1}{2}}=\frac{\sqrt{2}}{\sqrt{P_p\beta_c\alpha_c(2-\alpha_c)}}
\begin{bmatrix}
1&-\cos2\theta\\
-\cos2\theta&1
\end{bmatrix}^{-\frac{1}{2}}\label{rqsquare} \; .
\end{equation}
}

The determinant of $\begin{bmatrix}
1&-\cos2\theta\\
-\cos2\theta&1
\end{bmatrix}$ is $1-\cos^2 2\theta=\sin^2 2\theta$, and thus is always non-negative, and non-zero for $\theta\neq 90^{\circ}$. 
Thus $\mathbb{R}_{\mathbf{q}_{c,j}}$ is non-singular, positive definite and invertible. 
As a result, the probability expression in (\ref{cconstraint}) can be equivalently written as
\begin{align}
\mathbb{P}\{\mathbf{w}_{c,j}(\check{\mathbf{x}}_c)\geq\mathbf{q}_{c,j}\}&=\mathbb{P}\{\mathbf{w}_{c,j}(\check{\mathbf{x}}_c)\geq\mathbb{R}_{\mathbf{q}_{c,j}}^{\frac{1}{2}}\mathbb{R}_{\mathbf{q}_{c,j}}^{-\frac{1}{2}}\mathbf{q}_{c,j}\}\\
&=\mathbb{P}\{\mathbb{R}_{\mathbf{q}_{c,j}}^{-\frac{1}{2}}\mathbf{w}_{c,j}(\check{\mathbf{x}}_c)\geq\mathbb{R}_{\mathbf{q}_{c,j}}^{-\frac{1}{2}}\mathbf{q}_{c,j}\}\\
&=\mathbb{P}\{\bar{\mathbf{w}}_{c,j}(\check{\mathbf{x}}_c)\geq\bar{\mathbf{q}}_{c,j}\}\label{ppbar}
\end{align}
where $\bar{\mathbf{w}}_{c,j}(\check{\mathbf{x}}_c)\triangleq\mathbb{R}_{\mathbf{q}_{c,j}}^{-\frac{1}{2}}\mathbf{w}_{c,j}(\check{\mathbf{x}}_c)$ and $\bar{\mathbf{q}}_{c,j}\triangleq\mathbb{R}_{\mathbf{q}_{c,j}}^{-\frac{1}{2}}\mathbf{q}_{c,j}$. 
Consequently, the chance constraint (\ref{cconstraint}) is equivalent to 
\begin{equation}
\mathbb{P}\{\bar{\mathbf{w}}_{c,j}(\check{\mathbf{x}}_c)\geq\bar{\mathbf{q}}_{c,j}\}\geq v_c\label{ppbarv}
\end{equation}
with $\bar{\mathbf{q}}_{c,j}\sim \mathcal{N}(\mathbf{0},\; \mathbf{I})$.

To obtain an efficiently computable constraint, we apply the {\em Safe Approximation I} method in \cite{haqiqatnejad2020robust}.
The two entries of $\bar{\mathbf{q}}_{c,j}$ are uncorrelated and independent. Defining
\begin{equation}
\bar{\mathbf{q}}_{c,j}\triangleq\begin{bmatrix}\bar{q}_{c,j}^1\\\bar{q}_{c,j}^2\end{bmatrix},
\quad
\bar{\mathbf{w}}_{c,j}(\check{\mathbf{x}}_c)\triangleq\begin{bmatrix}\bar{w}_{c,j}^1\\\bar{w}_{c,j}^2\end{bmatrix} \; ,
\end{equation}
the Gaussian cumulative distribution function can be used to calculate the joint probability in~(\ref{ppbar}) as follows:
\begin{align}
\mathbb{P}\{\bar{\mathbf{w}}_{c,j}(\check{\mathbf{x}}_c)\geq\bar{\mathbf{q}}_{c,j}\}&=\mathbb{P}\{\bar{w}_{c,j}^1\geq\bar{q}_{c,j}^1\}\mathbb{P}\{\bar{w}_{c,j}^2\geq\bar{q}_{c,j}^2\}\\
&=\frac{1+\erf(\frac{\bar{w}_{c,j}^1}{\sqrt{2}})}{2}\times \frac{1+\erf(\frac{\bar{w}_{c,j}^2}{\sqrt{2}})}{2} \; ,
\end{align}
where the error function is given by $\erf(x)=\frac{2}{\sqrt{\pi}}\int_{0}^{x}\exp(-t^2)\mathop{dt}$. Due to the monotonicity of $\erf(x)$, the desired probability is bounded below by
\begin{equation}
\mathbb{P}\{\bar{\mathbf{w}}_{c,j}(\check{\mathbf{x}}_c)\geq\bar{\mathbf{q}}_{c,j}\}\geq\left[\frac{1+\erf(\frac{\min\{\bar{w}_{c,j}^1,\bar{w}_{c,j}^2\}}{\sqrt{2}})}{2}\right]^2 \; .
\end{equation}
In order to satisfy the chance constraint (\ref{ppbarv}), it is sufficient to consider the deterministic constraint
\begin{equation}
\left[\frac{1+\erf\left(\frac{\min\{\bar{w}_{c,j}^1,\bar{w}_{c,j}^2\}}{\sqrt{2}}\right)}{2}\right]^2\geq v_c.
\end{equation}
Since $\bar{\mathbf{w}}_{c,j}(\check{\mathbf{x}}_c)\triangleq\mathbb{R}_{\mathbf{q}_{c,j}}^{-\frac{1}{2}}\mathbf{w}_{c,j}(\check{\mathbf{x}}_c)$, the constraint can be rewritten as
\begin{equation}
{\color{black}
\mathbb{R}_{\mathbf{q}_{c,j}}^{-\frac{1}{2}}\mathbf{w}_{c,j}(\check{\mathbf{x}}_c)\geq\sqrt{2}\erf^{-1}\left(2\sqrt{v_c}-1\right)\mathbf{1}_2\tag{C4}\label{C4}}
\end{equation}
where $\erf^{-1}(\cdot)$ denotes the inverse error function. We thus finally arrive at the following linear inequality constraint:
{\color{black}
\begin{equation}
\widetilde{\mathbf{H}}_{cc,j}^{\mathrm{\mho}}\check{\mathbf{x}}_c\geq\bar{\alpha_c}\eta_c\mathbb{R}_{\mathbf{q}_{c,j}}^{\frac{1}{2}}\mathbf{1}_2 +{\color{black}\delta_{c,j}^0}\mathbf{1}_2
\end{equation}
}
where $\eta_c\triangleq \sqrt{2}\erf^{-1}\left(2\sqrt{v_c}-1\right)$.

\bibliographystyle{IEEEtran}
\bibliography{IEEEabrv,RobustRef}
\end{document}